\documentclass[aps,showpacs,showkeys,nofootinbib,twocolumn]{revtex4}
\usepackage{epsfig,amssymb,bm,graphicx,color,amsmath,rotating}
\usepackage{xcolor}
\usepackage[percent]{overpic}
\usepackage{mwe,xcolor}
\usepackage{transparent}
\usepackage{framed}
\usepackage{slashed}
\usepackage{mathtools}

\newcommand*{\ee}{e^+e^-}

\begin{document} %{\normalsize }
%\interfootnotelinepenalty=10000

\title{Positron energy distribution in factorized trident process
%at LUXE experiment
}

\author{A.~I.~Titov${}^1$, U. Hernandez Acosta${}^{2, 3}$, B.~K\"ampfer${}^{2, 4}$}

\affiliation{${}^1$Bogoliubov Laboratory of Theoretical Physics,
  JINR, Dubna 141980, Russia}
\affiliation{${}^2$Helmholtz-Zentrum  Dresden-Rossendorf, 01314 Dresden,
  Germany}
\affiliation{${}^3$Center for Advanced Systems Understanding, 
Helmholtz-Zentrum Dresden-Rossendorf e.V. (HZDR)
Untermarkt 20, 02826 Görlitz, Germany}
\affiliation{${}^4$Institut f\"ur Theoretische Physik, TU~Dresden, 01062
Dresden, Germany}

\begin{abstract}
  We estimate the energy distribution of positrons produced in the interaction
  of ultra-relativistic electrons with a high-intensity laser beam.
The underlying trident process is factorized on the probabilistic level.
That is, we deploy a two-step mechanism for the formation of electron-positron
  pairs. In the first step,
  a high-energy photon is produced as a result of nonlinear
  Compton scattering. In the second step, an electron-positron pair
  is created by the nonlinear (multi-photon) Breit-Wheeler process.
%  We provide the spectra
 % of positrons depending on the energy of the incident electron beam
%  and the intensity of the laser.
\end{abstract}

  
\date{\today}

\maketitle

\section{Introduction}

Our study is devoted to an analysis of the energy spectra
of positrons ($e^+$) produced in collisions 
of ultra-relativistic electrons ($e^-$) with a high-intensity laser beam
(${\cal L}$, circular polarization),
$e^- + {\cal L}\to {e^-}' + \ee$,
where ${e^-}'$ is the recoil electron. 
This is the trident process which, analog to the nonlinear two-photon Compton
process \cite{Loetstedt:2009zz,Seipt:2012tn,Mackenroth:2012rb}, 
is described within the Furry picture by a prototypical 
two-vertex diagram and its exchange part. Quantum interference effects
make a throughout description quite challenging. Correspondingly, a series
of papers, e.g.\  
\cite{Ilderton:2010wr,Hu:2010ye,King:2013osa,King:2013zw,Dinu:2017uoj,Torgrimsson:2020wlz,Dinu:2019wdw,Dinu:2019pau,Torgrimsson:2020gws,King:2018ibi,Mackenroth:2018smh},
analyzed in depth the many interesting facets of trident 
\cite{VIRitus,VNBaier}
as a fundamental strong-field QED process.
Trident is to be considered as the fist stage of the development of 
QED cascades and plays thus a prominent role, also for testing numerical
simulation codes \cite{Gonoskov:2014mda,DiPiazza:2018bfu,Gonoskov:2021hwf}, 
in particular in the strong-field regime.
The increasing interest is documented by a number of plans of experimental
facilities \cite{E_320,Meuren:2020nbw,Abramowicz:2019gvx,Abramowicz:2021zja,Salgado:2021fgt}
enabling high-precision measurements after the pioneering work
of E-144 \cite{Bamber:1999zt,Burke:1997ew} and successors \cite{Cole:2017zca,Poder:2018ifi}. 

In view of these developments, an easily usable approximation of trident 
should be devised. Following suggestions in \cite{Torgrimsson:2020wlz,Dinu:2019pau,Dinu:2019wdw} 
and in line with earlier approaches \cite{King:2013osa,King:2013zw,Hu:2010ye,Bamber:1999zt}, 
we apply a folding model of type
n$\ell$C $\otimes$ n$\ell$BW, i.e.\ gluing the probabilities of
nonlinear Compton (n$\ell$C)
and nonlinear Breit-Wheeler (n$\ell$BW)
processes as depicted by the Furry picture diagrams\\
\setlength{\unitlength}{1mm}
\thicklines
\hspace*{1mm}
\begin{picture}(8,9)
\put(0,0.5){\line(1,0){8}}
\put(0,1.0){\line(1,0){8}}
\put(3.0,0.5){\line(0,1){5.5}}
\put(3.0,5.5){\line(1,0){5}}
\put(3.0,5.9){\line(1,0){5}}
\put(3.0,5.5){\line(2,-1){5}}
\put(3.0,4.8){\line(2,-1){5}}
\put(9,4.7){{\footnotesize $e^+$}}
\put(9,2.2){{\footnotesize $e^-$}}
\put(9,0.0){{\footnotesize $e'$}}
\put(3,0.5){\circle*{1}}
\put(3,5.3){\circle*{1}} 
\end{picture}
\begin{picture}(8,8)
\put(4.9,3.0){$-$}
%\put(0,2.0){$-$}
\end{picture}
%$-$
\begin{picture}(8,9)
\put(-20.1,0){\line(0,1){8}}
\put(0,0.5){\line(1,0){8}}
\put(0,1.0){\line(1,0){8}}
\put(3.0,0.5){\line(0,1){5.5}}
\put(3.0,5.5){\line(1,0){5}}
\put(3.0,5.9){\line(1,0){5}}
\put(3.0,5.5){\line(2,-1){5}}
\put(3.0,4.8){\line(2,-1){5}}
\put(9,4.7){{\footnotesize $e^+$}}
\put(9,2.2){{\footnotesize $e'$}}
\put(9,0.0){{\footnotesize $e^-$}}
\put(3.1,0.5){\circle*{1}} 
\put(3.1,5.3){\circle*{1}} 
\put(12.9,0){\line(0,1){8}}
\put(13.5,6){2}
\put(15,3){$\to$}
\put(20,0){\line(0,1){8}}
\put(21,0.5){\line(1,0){8}}
\put(21,1.0){\line(1,0){8}}
\put(25,0.5){\line(0,1){5.5}}
\put(25,0.5){\circle*{1}} 
\put(26,5.5){{\footnotesize $\omega'$}}
\put(30,0){\line(0,1){8}}
\put(30.75,6){2}
\put(32.5,3){$\otimes$}
\put(36.5,0){\line(0,1){8}}
\put(38,0.5){\line(0,1){5.5}}
\put(38.2,5.5){\line(1,0){5.2}}
\put(38.2,5.9){\line(1,0){5.2}}
\put(38.5,5.5){\line(2,-1){5}}
\put(38.5,4.8){\line(2,-1){5}}
\put(44.5,4.7){{\footnotesize $e^+$}}
\put(44.5,2.2){{\footnotesize $e^-$}}
\put(38.,5.3){\circle*{1}} 
\put(39,0){{\footnotesize $\omega'$}}
\put(49,0){\line(0,1){8}}
\put(49.5,6){2}
\end{picture} 
\setlength{\unitlength}{1pt}
\hspace*{44mm}
(double lines: Volkov wave functions, vertical lines: photon propagator 
or real unpolarized photon with frequency $\omega'$).
This is the incoherent product of two one-vertex processes
in the spirit of kinetic simulations, cf.\  \cite{Elkina:2010up,Seipt:2020uxv}, 
%and can be validated by comparison with forthcoming data.
which is realized here by the convolution
\begin{equation}\label{convolution}
\int d \omega' \frac{d\Gamma_{n\ell C}^{e^- + {\cal L} \to \gamma'(\omega') + {e^-}'}}{d\omega'} \,
\Gamma_{n\ell BW}^{\gamma'(\omega') + {\cal L} \to e^+ e^-}  
\end{equation}
of  n$\ell$C and n$\ell$BW probabilities or rates $\Gamma_{n\ell C, n\ell BW}$
(see also equation (21) in \cite{King:2013osa} and equation (30) in \cite{King:2013zw}).
The integration over the energy $\omega'$ of the intermediate photon
reduces strongly the impact of $\omega'$-local structures of the individual rates. 
The recent references \cite{King:2019igt,King:2020hsk} discuss various
approximation schemes of one-vertex processes and their capabilities to catch the
rich structures in laser pulses. The convolution (\ref{convolution}) make them
less severe.

Our focus is on the positron energy distribution. 
The pair production probabilities have been
considered in \cite{Acosta21} within the same folding model.
Positrons are best identifiable and do not suffer from distinguishing the
recoil electron from the pair electron. The selected kinematics is motivated
by plans of the LUXE collaboration \cite{Abramowicz:2019gvx,Abramowicz:2021zja}.   
The considered laser intensities are high but not ultra-high, so that we are
working in a regime where the Ritus-Naroshny conjecture does not apply, 
therefore enabling the safe use of the Furry picture.
The two building blocks of our approach, i.e.\ n$\ell$C and n$\ell$BW,
have been elaborated long ago by Nikishov and Ritus, see \cite{Ritus85}.     

Our short note is organized as follows.
In section~\ref{NL-largexi} we recall the main formulas for 
nonlinear Compton and Breit-Wheeler processes
w.r.t.\ to the easy application in calculating 
the energy distribution of the produced positrons in electron-laser
collisions which we consider in section~\ref{eL2pair}.
Besides that also the laser intensity dependence is discussed
and a brief remark on the ``boiling point of the QED vacuum" is added there.
Our summary and remarks for further developments 
are given in section~\ref{summary}.
Band width effects are considered in appendix \ref{Appendix_A} 
for linear trident.

\section{Nonlinear one-vertex processes 
%at large \boldmath{$\xi$}
}\label{NL-largexi}

Analog to kinetic theory on probabilistic level as causal reaction network 
we approximate the factorized trident process by multiple emission 
as incoherent combination of one-vertex processes. Albeit nonlinear Compton
and nonlinear Breit-Wheeler processes are related by crossing symmetry 
on amplitude level, their final phase spaces are different and 
the physically allowed regions in the Mandelstam plane have different shapes,
as exhibited in figure 1 in \cite{HernandezAcosta:2020agu}. Our ansatz
in section \ref{eL2pair} below, based on the convolution (\ref{convolution}), needs the rates  
$\frac{d\Gamma_{n\ell C}(\omega' \vert E_e)}{d\omega'}$ and
$\frac{d\Gamma_{n\ell BW}(E^+ \vert \omega')}{dE^+}$   
which depend parametrically on laser frequency and intensity. 
The lab.\ frame 
kinematic quantities $E_e$ ($in$-electron energy), $\omega'$ (hard-photon energy)
and $E^+$ ($out$-positron energy)
uncover certain ranges: $0 < \omega' < E_e$  (n$\ell$C)
and $0 < E^+ < \omega'$ (n$\ell$BW),
suggesting representations of $out$-states quantities scaled by $in$-state quantities,
i.e.\ $0 < \omega'/E_e < 1$ and $0 < E^+ / \omega' < 1$ and finally
$0 < E^+ / E_e < 1$. The different dependencies of n$\ell$C and n$\ell$BW
rates on $(out \vert in)$ quantities also enable different approximations
which we recollect in the subsequent subsections on the basis of 
adapted textbook formulas.
 
For the classical laser intensity parameter we use the 
Lorentz and gauge invariant variable 
$\xi =  \vert e \vert {\cal E} /(m \omega)$,  expressed
by quantities in the lab.\ system:
${\cal E}$ - electric laser field strength (peak value),
$\omega$ - the laser's central  frequency; 
$- \vert e \vert$
and $m$ stand for the electron charge and mass, respectively,
and
$\alpha = e^2 / 4 \pi \approx 1/137$ is the 
fine-structure constant.
We use natural units with $\hbar = c = 1$.

\subsection{Nonlinear Compton scattering}

The following notation is employed for the nonlinear Compton process
$e + {\cal L} \to e' + \gamma'$:
$p$ is the $in$-electron four-momentum,
$k$ the laser photon four-momentum, 
and $k'$ the $out$-photon's ($\gamma'$) four-momentum. 
Quantities in the lab.\ frame
$E_e$, $\omega$ and $\omega'$ denoting the incoming (beam) electron energy
mentioned above,
frequency of the above mentioned laser beam 
and Compton-produced photons, respectively.
For simplicity, we restrict ourselves on head-on electron-laser collisions
and leave the explication of formulas for the experimentally relevant case
of a finite inclination of both beams to later work.
The laser frequency is chosen as $\omega = 1.55$~eV.

The one-photon emission rate in a monochromatic,
circularly polarized background (laser) field 
is \cite{Ritus85,Kampfer:2020cbx,Acosta21}
\begin{eqnarray} \label{nlCo}
\frac{d\Gamma_{n\ell C}}{d\omega'} &=& \frac{\alpha m^2}{E_e^2}
\sum_{n=n_{min}}^\infty 
{\cal F}_n (z, u, \xi), \\ 
 \label{nlCo_Bessel}
{\cal F}_n = - J_n^2(z) &+& \xi^2 w(u)
\left( \left[\frac{n^2}{z^2} - 1 \right] J_n^2(z) + {J_n'}^2(z) \right), %\nonumber
\end{eqnarray}
where $w(u) = 1 + \frac{u^2}{2 (1 + u)}$ and 
\begin{eqnarray}
&&n_{min} = \left\lceil 1 + \frac{m^2 (1 + \xi^2) \omega'}{4 \omega E_e (E_e - \omega')} \right\rceil , \\
&&z = \frac{\xi^2 \sqrt{1 + \xi^2}}{\chi} \sqrt{u (u_n - u)}, \quad \chi = \xi k\cdot p /m^2,\\
&&u=k \cdot k'/ k \cdot p' \approx \omega'/(E_e-\omega'), \,\,
u_n = \frac{2 n \chi}{\xi (1+\xi^2)}.
\end{eqnarray}
$J_n$ and $J'_n$ denote Bessel function of first kind and
its derivative, respectively, and  $\left\lceil \cdot \right\rceil$ stands for the ceiling function.
Some approximation of (\ref{nlCo_Bessel})
is provided by utilizing an asymptotic expression of the Bessel function 
\begin{equation} \label{Bessel_approx}
J_n(z)\approx (2\pi n\tanh a)^{-1}{\rm e}^{2n(a-\tanh a)}
\end{equation}
with $\tanh a=\sqrt{1-z^2/n^2}$ valid at $\xi\gg1$.
In numerically evaluating the differential rate of n$\ell$C,
the ``large-$\xi$ approximation" \cite{Ritus85} is often convenient:
\begin{equation}  \label{nlCo_approx} 
\frac{d\Gamma_{n\ell C}}{d\omega'}
\approx - \frac{\alpha m^2}{\pi E_e^2}
\left\{ \int\limits_{z}^{\infty} dy \, \Phi(y) 
+ \frac{2}{z}   w(u) \Phi'(z) \right\},
%\nonumber
\end{equation}
where 
$\Phi(z)$ and $\Phi(z)'$ stand
for the Airy function and its derivative
with arguments $z = (u/\chi)^{2/3}$.

\begin{figure}[t!]
\includegraphics[width=0.66\columnwidth]{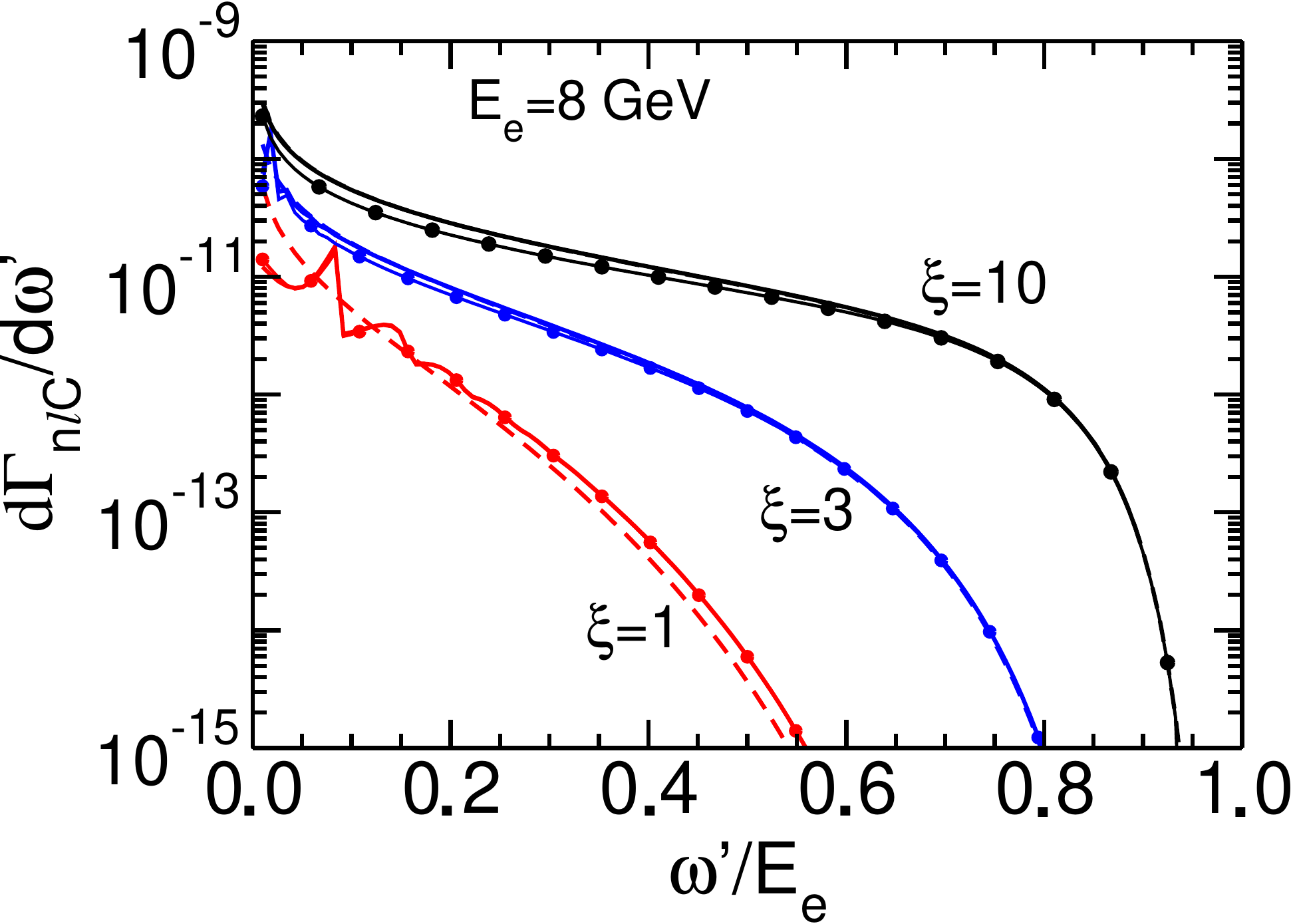}   
\includegraphics[width=0.66\columnwidth]{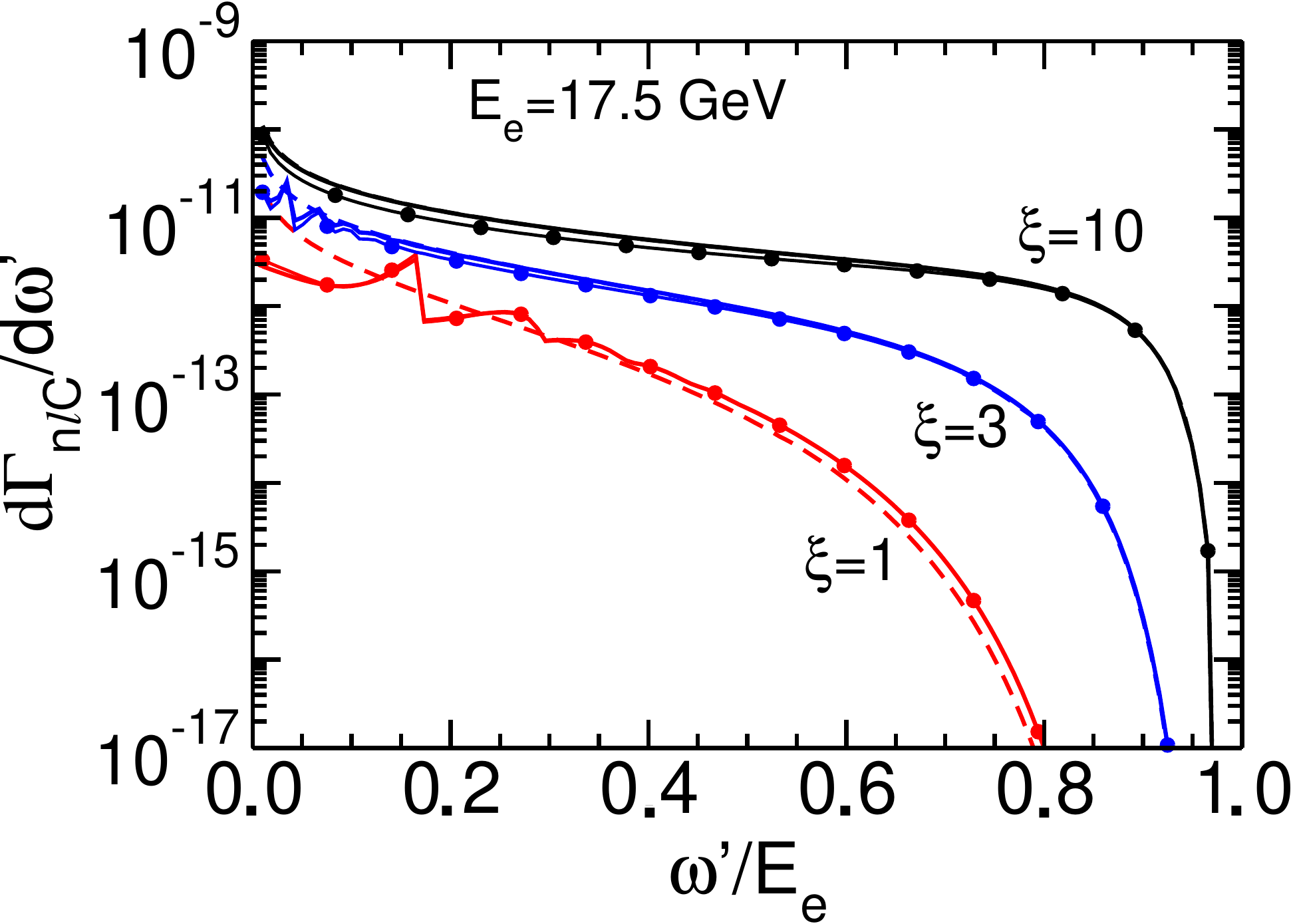}   
\includegraphics[width=0.66\columnwidth]{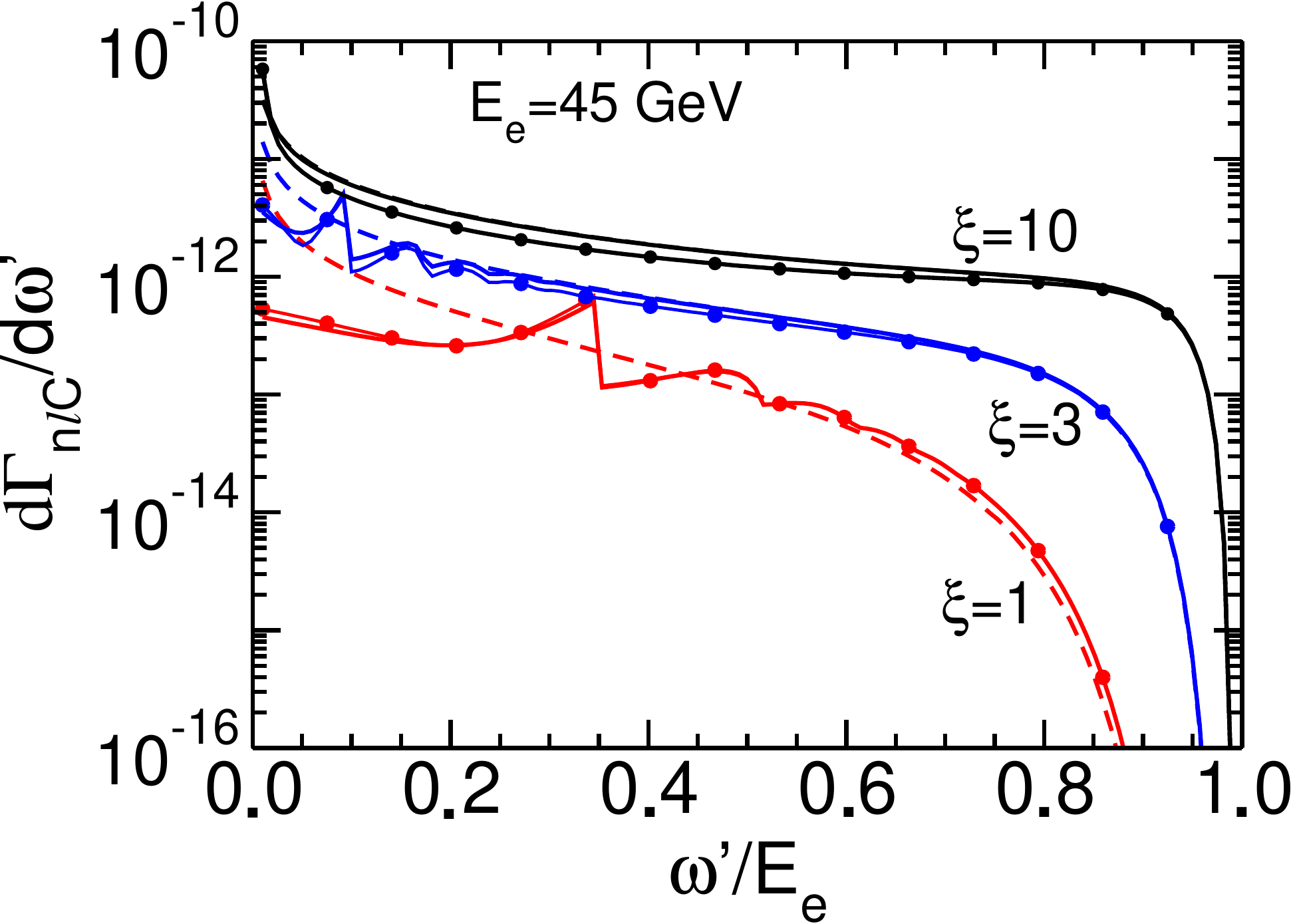}   
\caption{(Color online)
The dimensionless differential rate 
$d \Gamma_{n\ell C}/d\omega'$ as a function of
$\omega'$ (scaled by $E_e$) for electron beam
energies $E_e = 8$~GeV (top panel),
17.5~GeV (middle panel) and 45~GeV (bottom panel)
at different values of $\xi = 1$ (red),
3 (blue) and 10 (black).
The solid curves depict the  monochromatic model
Eqs.~(\ref{nlCo}, \ref{nlCo_Bessel}),
while dotted curves use the approximation Eq.~(\ref{Bessel_approx}).
Dashed curves are for the approximation Eq.~(\ref{nlCo_approx}).
The solid and dotted curves display harmonic structures
at low values of $\omega'$ and $\xi$.
\label{Fig:01}}
\end{figure}

The differential distributions $d\Gamma_{n\ell C}/d\omega'$ for various
initial electron energies $E_e$ and laser intensities $\xi$ 
are exhibited in Fig.~\ref{Fig:01}.\\
The solid curves depict the monochromatic
model, Eqs.~(\ref{nlCo}, \ref{nlCo_Bessel}),
and the dotted and dashed curves correspond to the approximations
(\ref{Bessel_approx}) and (\ref{nlCo_approx}), respectively.
Results depicted by solid and dotted curves are close to
each other in the considered region of $E_e$, $\xi$ and $\omega'$
At $\xi\geq3$ and $\omega'\geq 0.1E_e$, the
solid and dashed curves are indistinguishable.
One can see a monotonic decrease of the rate with increasing
$\omega'$ and some plateau at $\omega' \sim E_e/2$.
The width of the plateau increases with increasing $E_e$.
The harmonic structures of Eqs.~(\ref{nlCo}, \ref{nlCo_Bessel}) 
are visible at low values of $\omega'$ and extend
only for small values of $\xi$ towards moderate values of $\omega'$.
Despite the assumptions made in deriving
the approximation Eq.~(\ref{nlCo_approx}) \cite{Ritus85} the agreement with
Eq.~(\ref{nlCo_Bessel}) is impressive for hard photons at $\xi > 1$.

When considering a laser pulse, instead of the monochromatic background field,
an additional scale enters, which characterizes,
e.g.\ the pulse duration. Due to
interference effects, the $out$-photon spectra may acquire more complex shapes
with nonlinear dependencies on pulse shape and duration. There is extended
literature on this and attempts of useful approximation schemes.
We emphasize again that, in our positron number estimate below,
the rate $\frac{d\Gamma_{n\ell C}(\omega' \vert E_e)}{d\omega'}$ 
is to be integrated over $\omega'$ with the weight
$\frac{d\Gamma_{n\ell BW}(E^+ \vert \omega')}{dE^+}$
at given $E^+$. In so far, the (fine or harmonic) structures in 
$\frac{d\Gamma_{n\ell C}(\omega' \vert E_e)}{d\omega'}$ 
do not matter. The agreement of the differential cross section for a pulse
(see figure 8 in \cite{Kampfer:2020cbx}) 
with the monochromatic spectrum
according to Eqs.~(\ref{nlCo}, \ref{nlCo_Bessel})  
support that claim.
As a consequence,  Eqs.~(\ref{nlCo}, \ref{nlCo_Bessel}, \ref{Bessel_approx}, \ref{nlCo_approx}) provide
useful ingredients of our n$\ell$C$\otimes$n$\ell$BW folding model.

\subsection{Nonlinear Breit-Wheeler process}

The differential rate of produced positrons in
nonlinear (essentially multi-photon) Breit-Wheeler process
$\gamma' + {\cal L} \to e^+ e^-$
for given four-momentum $k'$ ($in$-photon $\gamma'$, frequency $\omega'$) and 
$k$ (laser, frequency $\omega$) reads for head-on collisions \cite{Ritus85}
\begin{eqnarray}  \label{nlBW}
\frac{d\Gamma_{n\ell BW}}{dE^+} &=&
\frac{\alpha m^2}{{\omega'}^2}
  \sum\limits_{n=n_{\rm min}}^{\infty}
 {\cal J}_n(z, u, \xi) , \\
%\end{equation} %\begin{equation}  
\label{nlBW_Bessel}
{\cal J}_n =
J_n^2(z) &+& \xi^2 w(u)
  \left(
  \left[\frac{n^2}{z^2}-1\right] J_n^2 (z)+{J_n'}^2 (z) \right),
\end{eqnarray} 
where $E^+$ is the positron energy, $w(u) = (2u-1)$ and
\begin{eqnarray}\label{variables}
n_{\rm min}&=&  \left\lceil 1+ \frac{m^2(1+\xi^2)\omega'}{4\omega E^+(\omega' - E^+)} \right\rceil~,
\\
z&=&\frac{2\xi n}{\sqrt{1+\xi^2}}
  \sqrt{\frac{u}{u_n} \left(1-\frac{u}{u_n} \right)}~, \label{eq:z} \\
u&=&\frac{{\omega'}^2}{4 E^+(\omega' -E^+)}~,
\qquad
u_n=\frac{n\omega\omega'}{m^2(1+\xi^2)}~. \label{eq:u_n}
\end{eqnarray}
Some approximation is based on Eq.~(\ref{nlBW_Bessel})
with (\ref{Bessel_approx}) yielding
\begin{equation} \label{nlBW_approx2}
{\cal J}_n(z, u, \xi) \approx \frac{1}{2 \pi}
{\rm e}^{2n(a-\tanh a)}
\frac{1+2 \xi^2(2u-1)\sinh^2a}{n\tanh a} . 
\end{equation}
Analog to Eq.~(\ref{nlCo_approx}) one can utilize the approximation \cite{Ritus85} 
\begin{equation}  \label{nlBW_approx1} 
\frac{d\Gamma_{n\ell BW}}{d E^+}
\approx  \frac{\alpha m^2}{\pi {\omega'}^2} 
\left\{ \int\limits_{z}^{\infty} dy \, \Phi(y)
- \frac{2}{z}(2u - 1) \Phi'(z) \right\} ,
\end{equation}
with $z = (4 u/ \chi_\gamma )^{2/3}$ and $\chi_\gamma = \xi k \cdot k'/m^2$,\
which both have been employed, e.g.\ in \cite{King:2013osa,King:2013zw}.

\begin{figure}[t!]
\includegraphics[width=0.66\columnwidth]{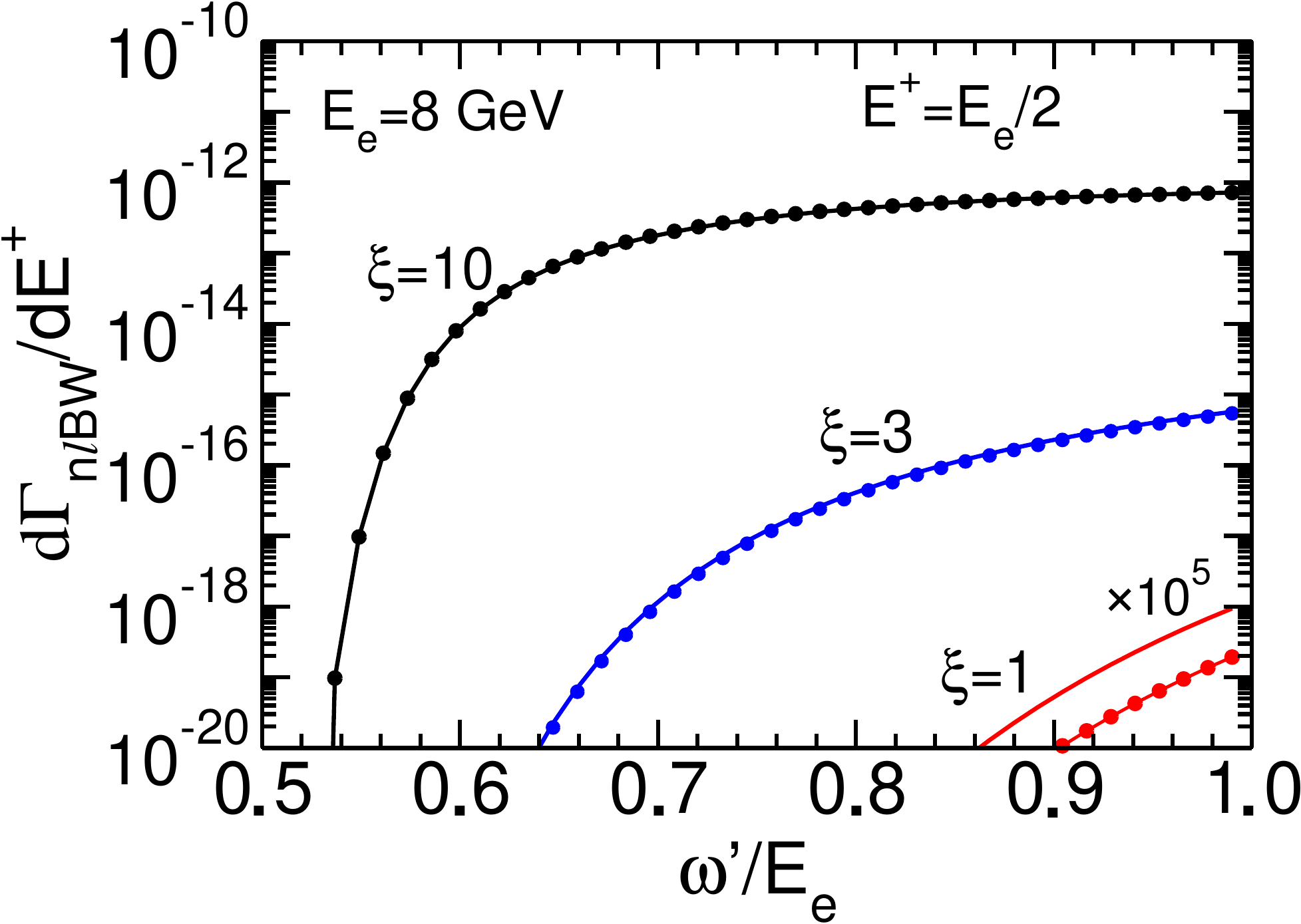}   
\includegraphics[width=0.66\columnwidth]{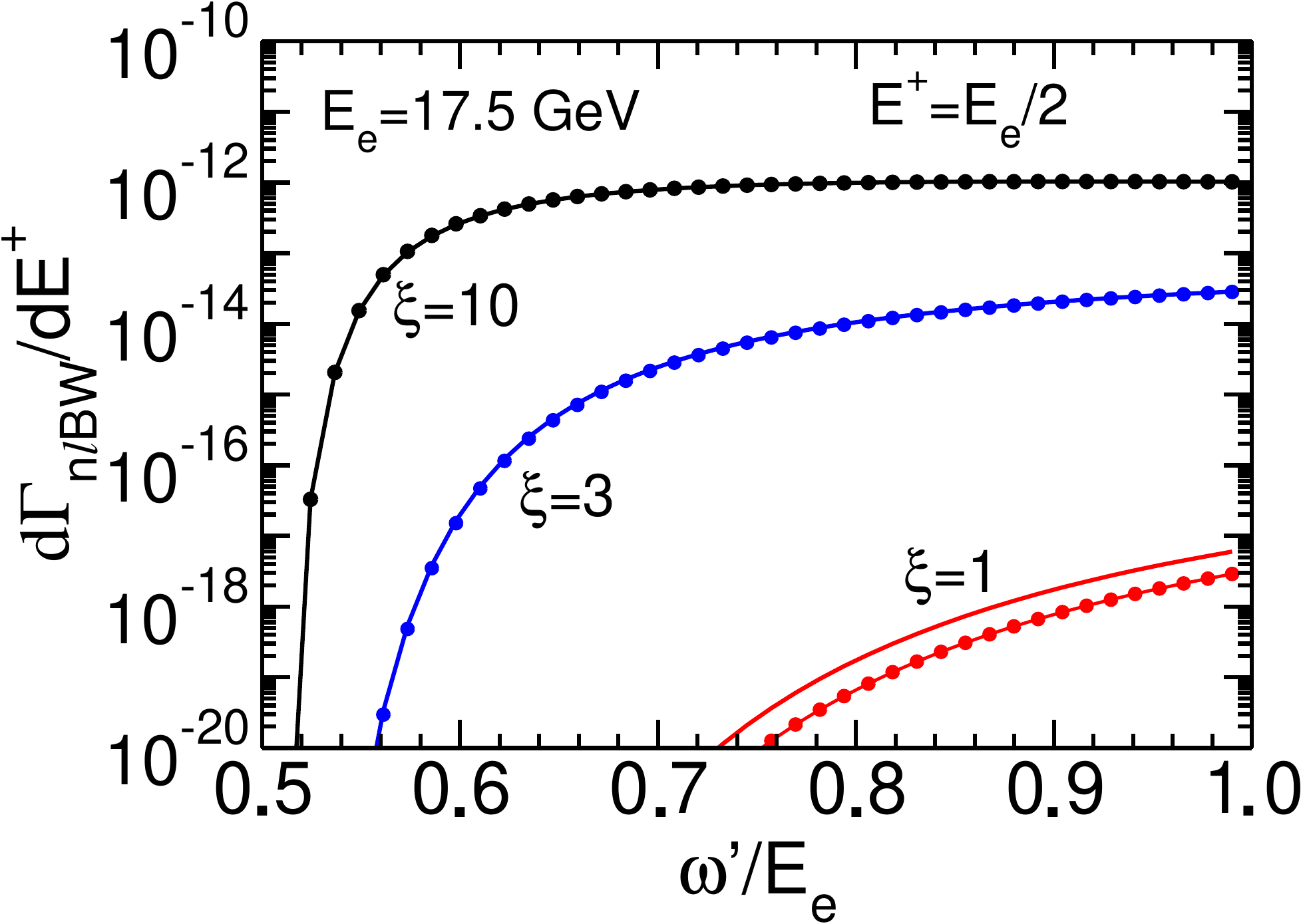}   
\includegraphics[width=0.66\columnwidth]{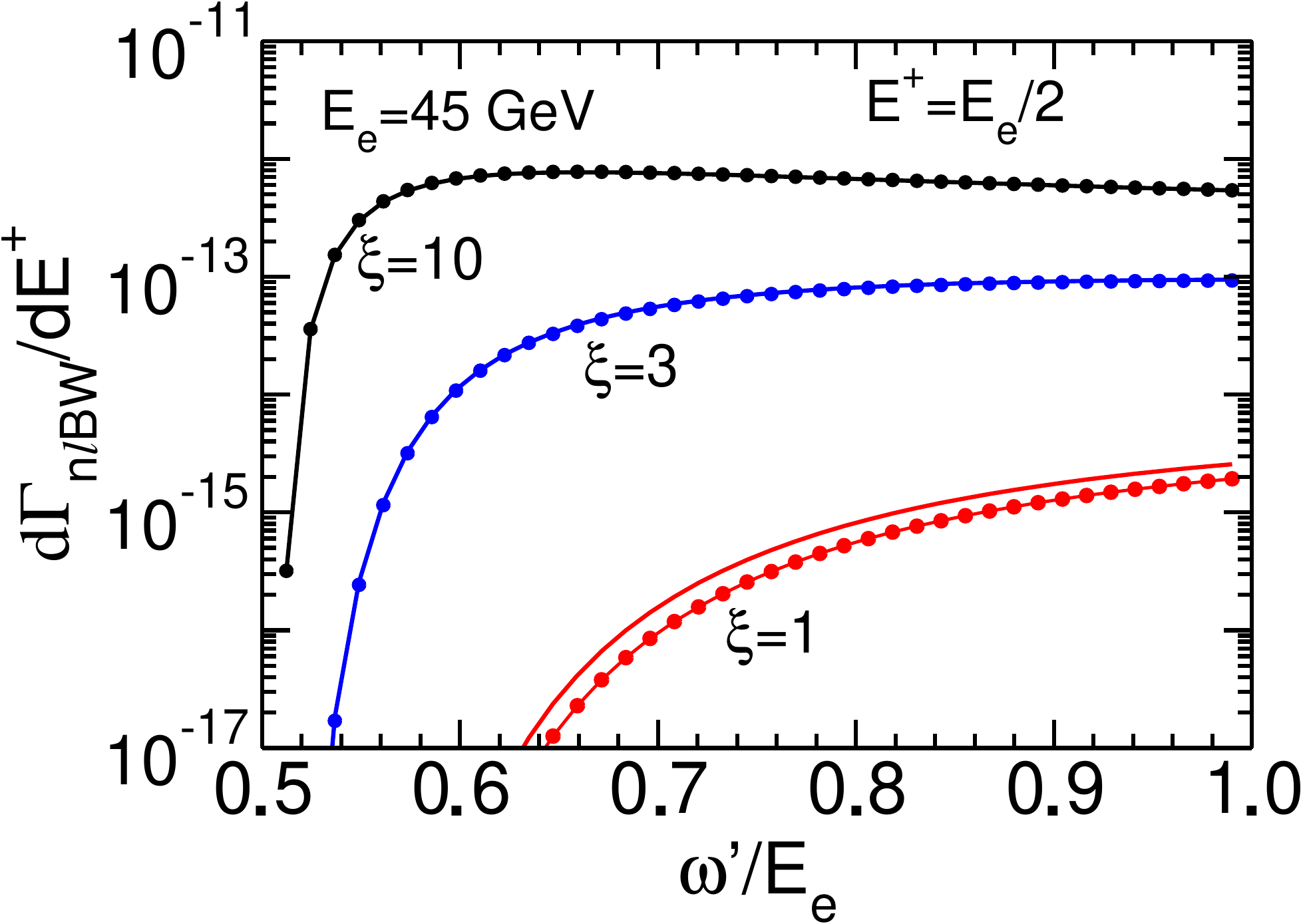}   
\caption{(Color online)
The dimensionless differential rate 
$d \Gamma_{n\ell BW}/dE^+$ as a function of
$\omega'$ (scaled by $E_e$) for positron energies $E^+ = E_e /2$ for
$E_e = 8$~GeV (top panel),
17.5~GeV (middle panel) and 45~GeV (bottom panel)
at different values of $\xi = 1$ (red), 3 (blue) and 10 (black).
Results using either Eqs.~(\ref{nlBW}, \ref{nlBW_Bessel}) (solid curves) or 
(\ref{nlBW}, \ref{nlBW_approx2}) (dashed curves) 
are not distinguishable on the given scale.
The approximation  (\ref{nlBW_approx1}) is displayed by dotted curves. 
\label{Fig:021}}
\end{figure}

The dependence of the rate $d \Gamma_{n\ell BW}/dE^+$ as a function of
$\omega'$ for fixed positron energy $E^+ = E_e /2$ is exhibited
in Fig.~\ref{Fig:021}.
Since $E^+ < \omega'$, only the range $0.5 < \omega'/E_e$ is accessible. 
The small (large)-$\omega'$ range becomes accessible
for small (large) values of $E^+$. 
Note the opposite tendencies of the Compton
and Breit-Wheeler rates as a function of $\omega'$,
making the details of small-$\omega'$ and
large-$\omega'$ distributions irrelevant
when considering their product in convoluting them.
The laser intensity dependency is also very strong
in the considered parameter range.

\begin{figure}[h!]
\includegraphics[width=0.49\columnwidth]{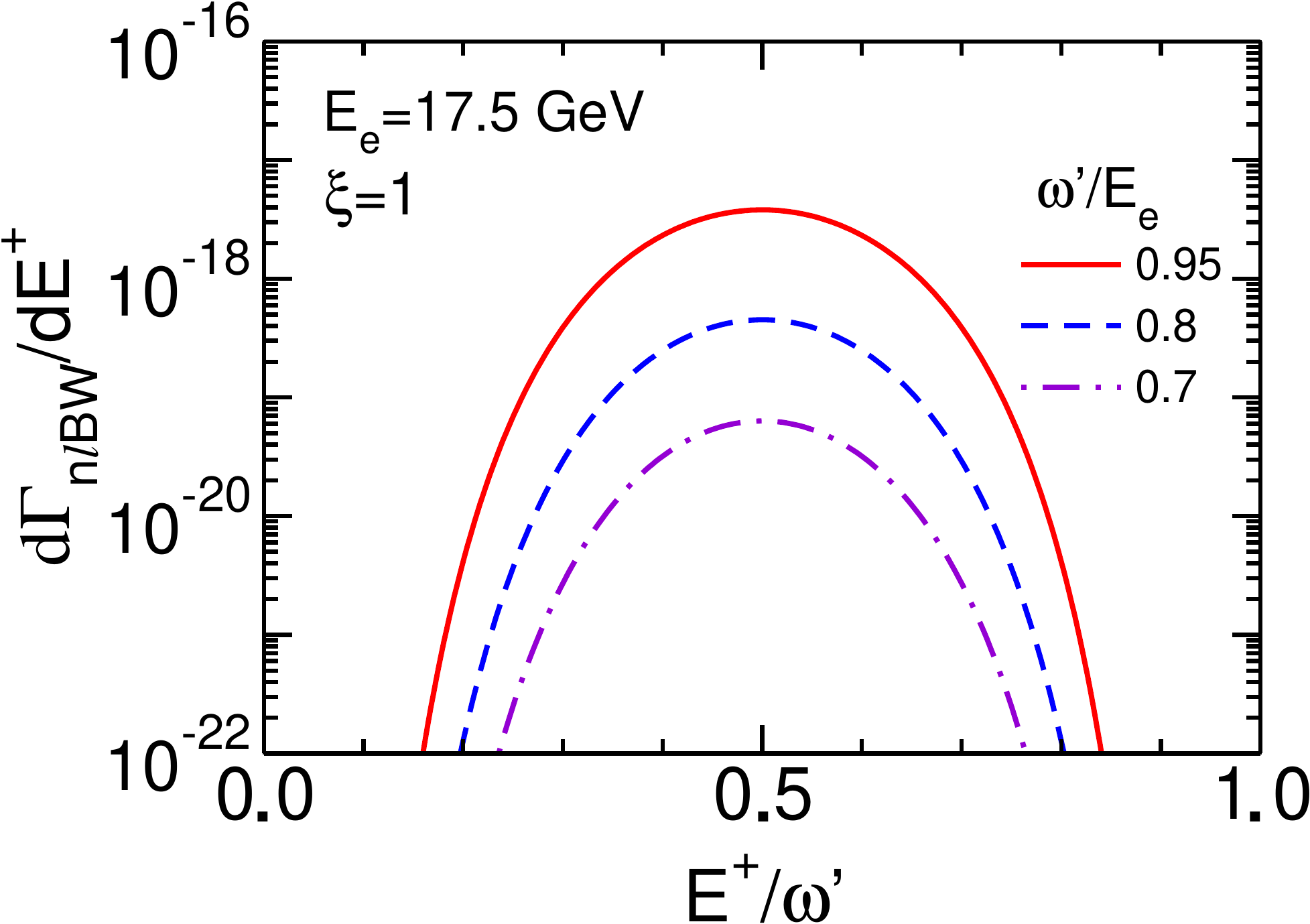} %\\
\includegraphics[width=0.49\columnwidth]{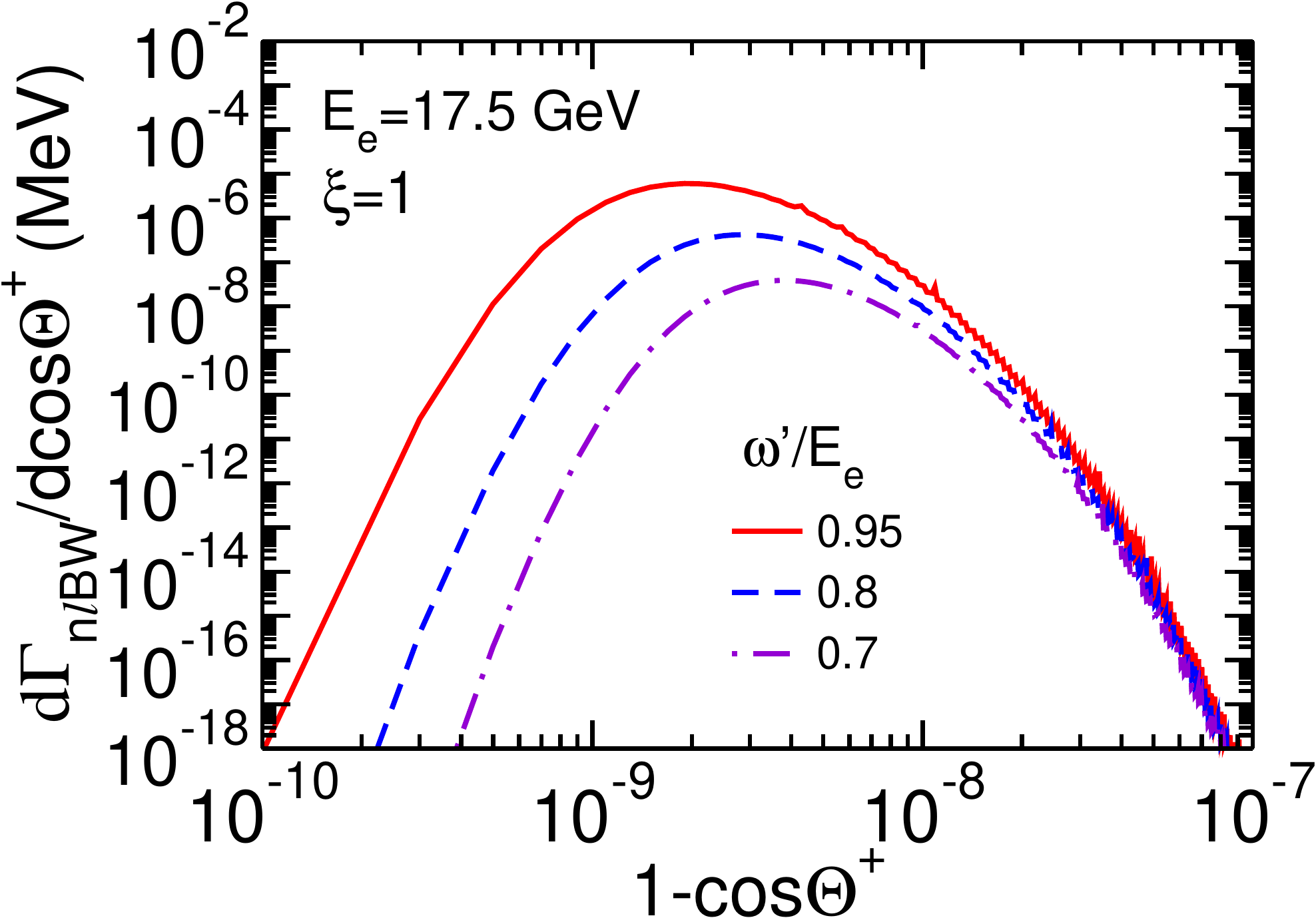}\\
\includegraphics[width=0.49\columnwidth]{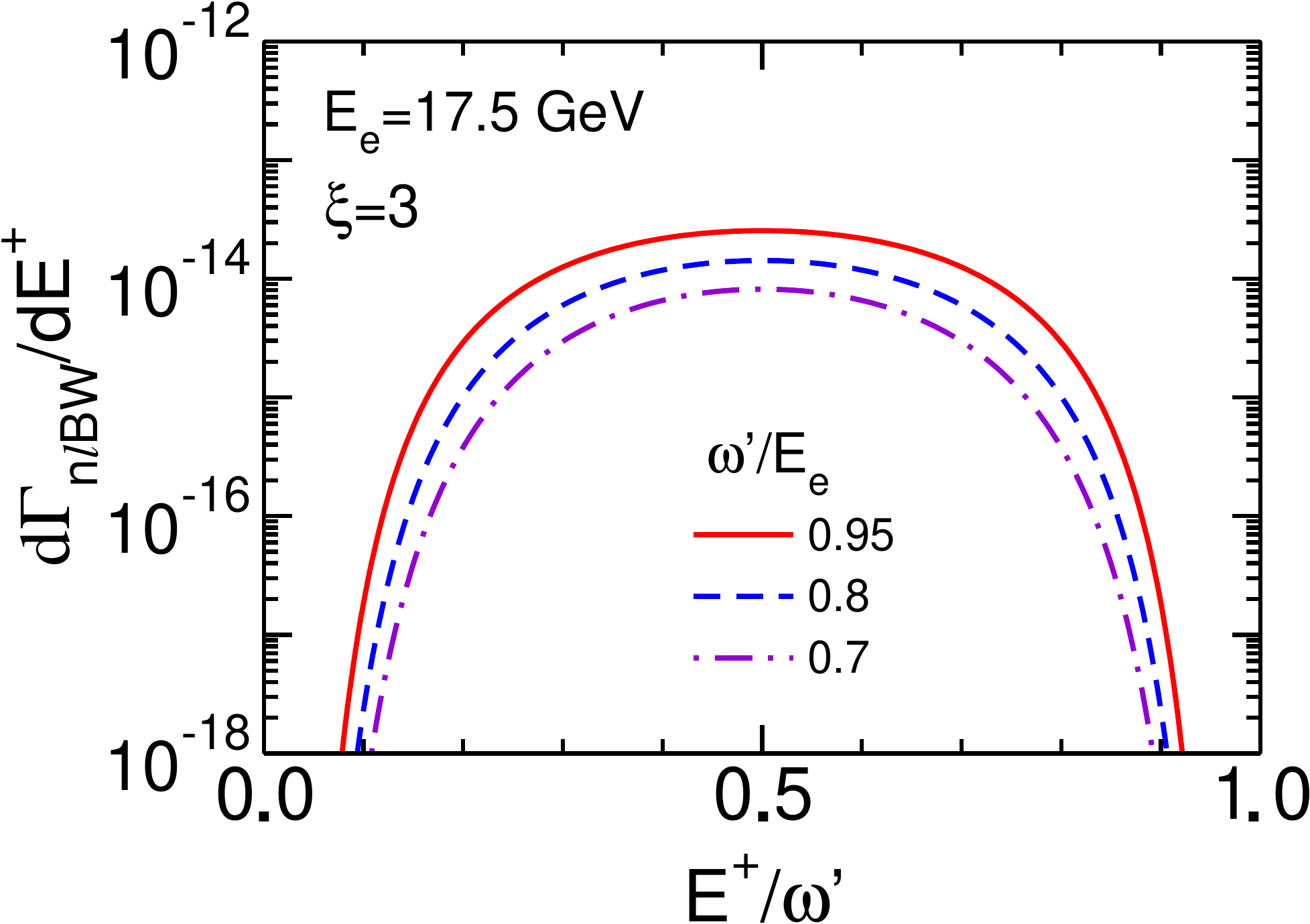} %\\
\includegraphics[width=0.49\columnwidth]{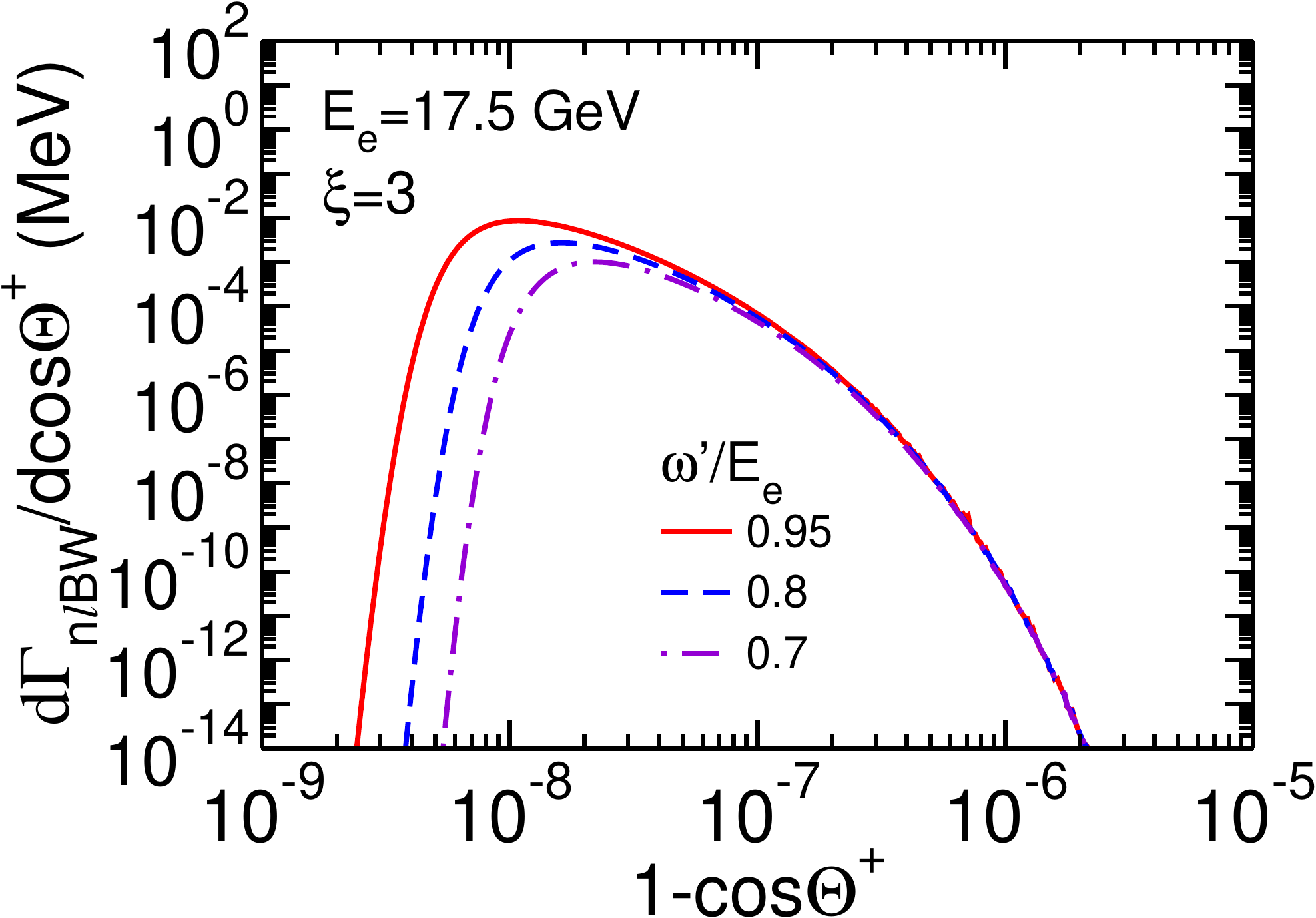}\\
\includegraphics[width=0.49\columnwidth]{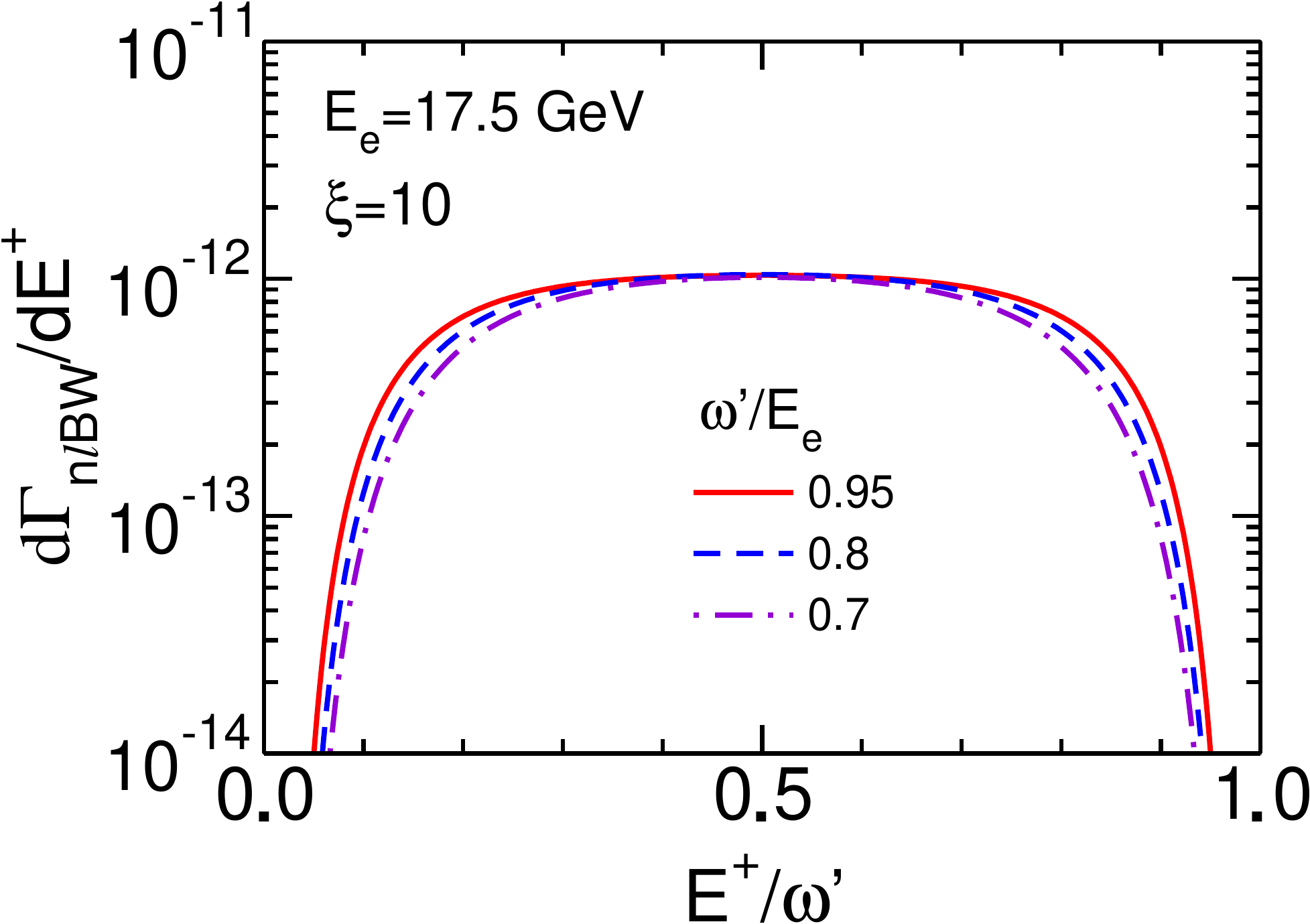}
\includegraphics[width=0.49\columnwidth]{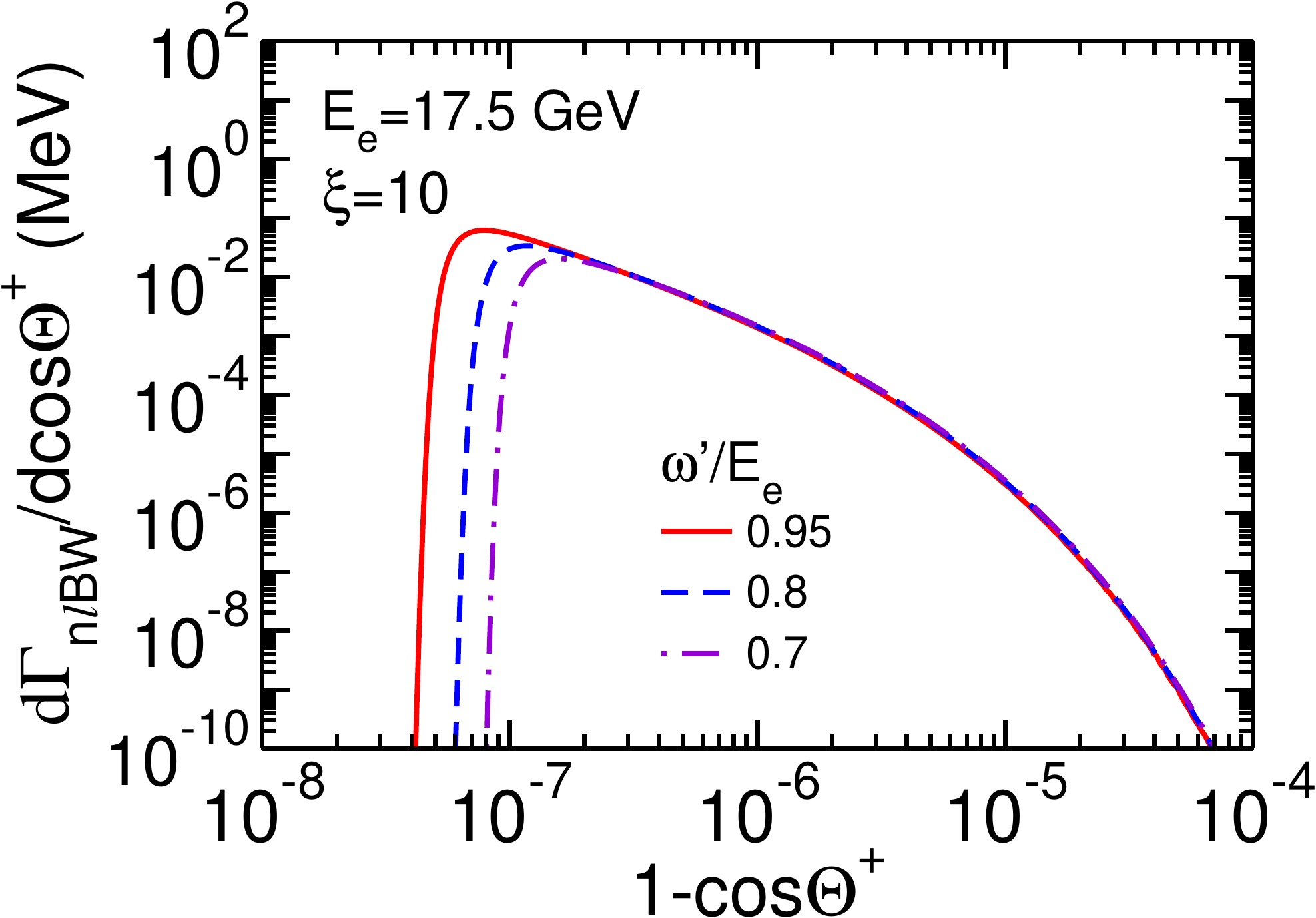}
\caption{(Color online)
Positron production rates $d\Gamma_{n\ell BW}/dE^+$ 
as a function of positron energy $E^+$ (scaled by $\omega'$)
(left row) and $d\Gamma_{n\ell BW}/d \cos \Theta^+$ as a function 
of the positron lab.\ angle $\Theta^+$ 
relative to the $\omega'$-photon direction 
(right row)
at laser intensity parameters $\xi = 1$ (top panel), 3 (middle panel) and
10 (bottom panel)
and various
energies of the probe photon $\omega'$ in units of $E_e=17.5$~GeV.
In calculation we use Eq.~(\ref{nlBW_approx2}).
\label{Fig:02}}
\end{figure}

Some information on the $E^+$ dependence of $d\Gamma_{n\ell BW}/dE^+$ 
at fixed values of $\omega'$
is given in the left column of Fig.~\ref{Fig:02} for
$E_e=17.5$~GeV at different values $\xi$.
The symmetric energy distributions
have a pronounced maximum at $E^+ = \omega'/2$
and monotonically decrease for $E^+ \to 0$ or $E^+ \to \omega'$,
thus evidencing the relevant positron energy range. 
To quantify the positron polar-angular distribution 
and to characterize the relevant range we exhibit these 
in the right column of Fig.~\ref{Fig:02}.
The angular distributions
exhibit a dead cone at small angles $\Theta'$ relative
to the $\omega'$ photon direction, i.e.\
at $(1 - \cos \Theta') \to 0$, similar to the
nonlinear Compton process \cite{HernandezAcosta:2020agu}. 
The dead cone effect becomes sharper with increasing value of $\xi$. 
The displayed angular-differential rate is determined by
\begin{equation}
\frac{d\Gamma_{n\ell BW}}{d \cos \Theta^+}
  =
\frac{\alpha m^2}{ \omega'}
  \sum\limits_{n=n_{\rm min}}^{\infty}
\frac{q^2}{\vert q \omega_+' - q_0 \omega_-' \cos \Theta^+ \vert}
 {\cal J}_n ,
\end{equation}  
where the definitions (\ref{eq:z}, \ref{eq:u_n}) apply for $z$ and $u_n$ 
entering ${\cal J}_n $ (\ref{nlBW_Bessel})
with $\omega_\pm' = \omega' \pm n \omega$, 
$q = \sqrt{q_0^2 - m^2 (1 + \xi^2)}$,
$q_0 = A + \sqrt{A^2 - B}$, as well as
\begin{eqnarray}
n_{min} &=&  \left\lceil 1 + \frac{ m^2 (1 + \xi^2)}{\omega \omega'}  \right\rceil, \\
u &=& \frac{{\omega'}^2} {(q_0 + q \cos \Theta^+)
(2 \omega' - q_0 - q \cos \Theta^+)},\\
A &=& \frac{2 n \omega \omega' \omega_+'}{{\omega_+'}^2 - {\omega_-'}^2
\cos^2 \Theta^+}, \\
B &=& \frac{m^2 (1 + \xi^2) {\omega_-'}^2 \cos^2 \Theta^+ 
+ 4 (n \omega \omega')^2}{ {\omega_+'}^2  - {\omega_-'}^2
\cos^2 \Theta^+}. 
\end{eqnarray}

\section{Estimate of positron production}\label{eL2pair}

The above recollected rates for monochromatic background (laser) fields
may be realized by a very long flattop section of the envelope. To be specific
consider a plane-wave laser model with electromagnetic vector potential 
$A (\phi) \propto  f(\phi) \, cos (\phi)$ with invariant phase $\phi = \omega (t + z)$,
i.e.\ a beam moving along $z$ axis to left, and lines of constant phase are
inclined by $45 ^{\circ}$ in the observer's space-time ($z$-$t$) diagram.
An idealized flattop envelope would be
$f(\phi) = \sqcap (\phi, \Delta \phi)$ with $\sqcap = 1$ for $\phi \in [0, \Delta \phi]$ and zero elsewhere,
i.e.\ the number of oscillations within the pulse is $N = \Delta \phi / 2 \pi$.
(Such an envelope has been considered in \cite{King:2020hsk} and \cite{Tang:2021qht}
for n$\ell$C and n$\ell$BW, respectively, with emphasis on the soft part
of the distributions and comparison of various approximation schemes.)
Despite intricate effects of finite-time laser pulses  \cite{King:2020hsk},
a simplified estimate of the number of produced quanta would be to use the 
monochromatic rate multiplied by the pulse duration.
In such a spirit we estimate the number of produced positrons by convoluting
the rate of produced hard photons, emitted by a high-energy electron -- traversing
a pulse on the trajectory $t \approx z$ near light cone thus crossing the
invariant phase lines $\phi = 0 \cdots \Delta \phi$ -- 
with the rate of positrons produced by these photons later on within the same pulse.
Let be the relevant time intervals $T_{C, BW}$ for the respective n$\ell$C and n$\ell$BW
processes. Then    
the energy distribution of positrons in the folding model
n$\ell$C $\otimes$ n$\ell$BW may be expressed as
\begin{equation}\label{E-distribution}
\frac{d N^{e^+}}{dE^+}
=\int\limits_0^{E_e} d\omega'
\int\limits_0^{T_C}dt \frac{d\Gamma_{n\ell C}(\omega',t)}{d\omega'} 
\int\limits_t^{T_{BW}}dt' \frac{d\Gamma_{n\ell BW}(\omega',t')}{dE^+} ,
\end{equation}
where $d\Gamma_{n\ell C}(\omega',t)/d\omega'$ is the 
rate of photons per frequency interval $d \omega'$ 
emerging from one electron undergoing Compton scattering
at time $t \in [0, T_C]$, and  
$d\Gamma_{n\ell BW}(\omega',t')/{dE^+}$ is the
the differential rate of Breit-Wheeler pairs
generated by a probe photon of frequency $\omega'$
at lab.\ frame time $t' \in [t, T_{BW}]$. 
Formally, $\Gamma_{n \ell C} (\omega', t) \to  \Gamma_{n \ell C} (\omega') \sqcap (t, T_C)$ and
$\Gamma_{n \ell BW} (\omega', t') \to  \Gamma_{n \ell BW} (\omega') \sqcap (t', T_{BW})$.
The above picture of an electron traversing
a laser pulse in head-on geometry near to light cone
means $T_C = T_{BW} \equiv T_0 = \pi N / \omega$.
(The pulse duration for an observer at rest in the $z$-$t$ Minkowski frame
is $2T_0$. On the trajectory $z \approx t$, the invariant phase is
$\phi \approx 2 \omega t$, therefore,
$\sqcap (t, T_0)  \to  \sqcap (\phi, \Delta \phi)$.) 
Neglecting spatio-temporal variations within the pulse
by using Eqs.~(\ref{nlCo}) and (\ref{nlBW}),
our final formula becomes 
\begin{eqnarray}\label{E-distribution-final}
%  {E-distribution-final}
\frac{d N^{e^+}}{dE^+}
=F_t\int\limits_{E^+}^{E_e} d\omega'
\frac{d\Gamma_{n\ell C}(\omega')}{d\omega'}\,
\frac{d\Gamma_{n\ell BW}(\omega',E^+)}{dE^+}~,
\end{eqnarray}
with $F_t =  N^e_0 \, T_C (T_{BW} - T_C / 2) = T_0^2 /2$, i.e.\
a quadratic dependence on the pulse length via $N^2$.
Such an estimate neglects the attenuation of the primary electron beam
towards emitting a hard photon at $t \approx z$ (which can be cured by
a Glauber eikonal factor), a formation time for separating the recoil electron
and the on-shell photon 
(which is here assumed to continue on the straight trajectory), 
and the poor approximation of the photon number
distribution in a finite pulse by
neglecting bandwidth effects (see Appendix \ref{Appendix_A}
for genuine bandwidth effects in the weak-field trident process).  
Obviously, the separation in one-step and the here only partially uncovered
two-step processes on probabilistic level is rather crude. Therefore,      
in our estimates we either use $T_0=2\pi/\omega$ 
(see Appendix B for a comparison with E-144 data)
and attach to the Compton rate the number $N^e_0 = 6 \times 10^9$ of
primary electrons per bunch 
or scale out the normalization factor $F_t$ \cite{footnote_in_bibliography}. 

\begin{figure}[t!]
\includegraphics[width=0.66\columnwidth]{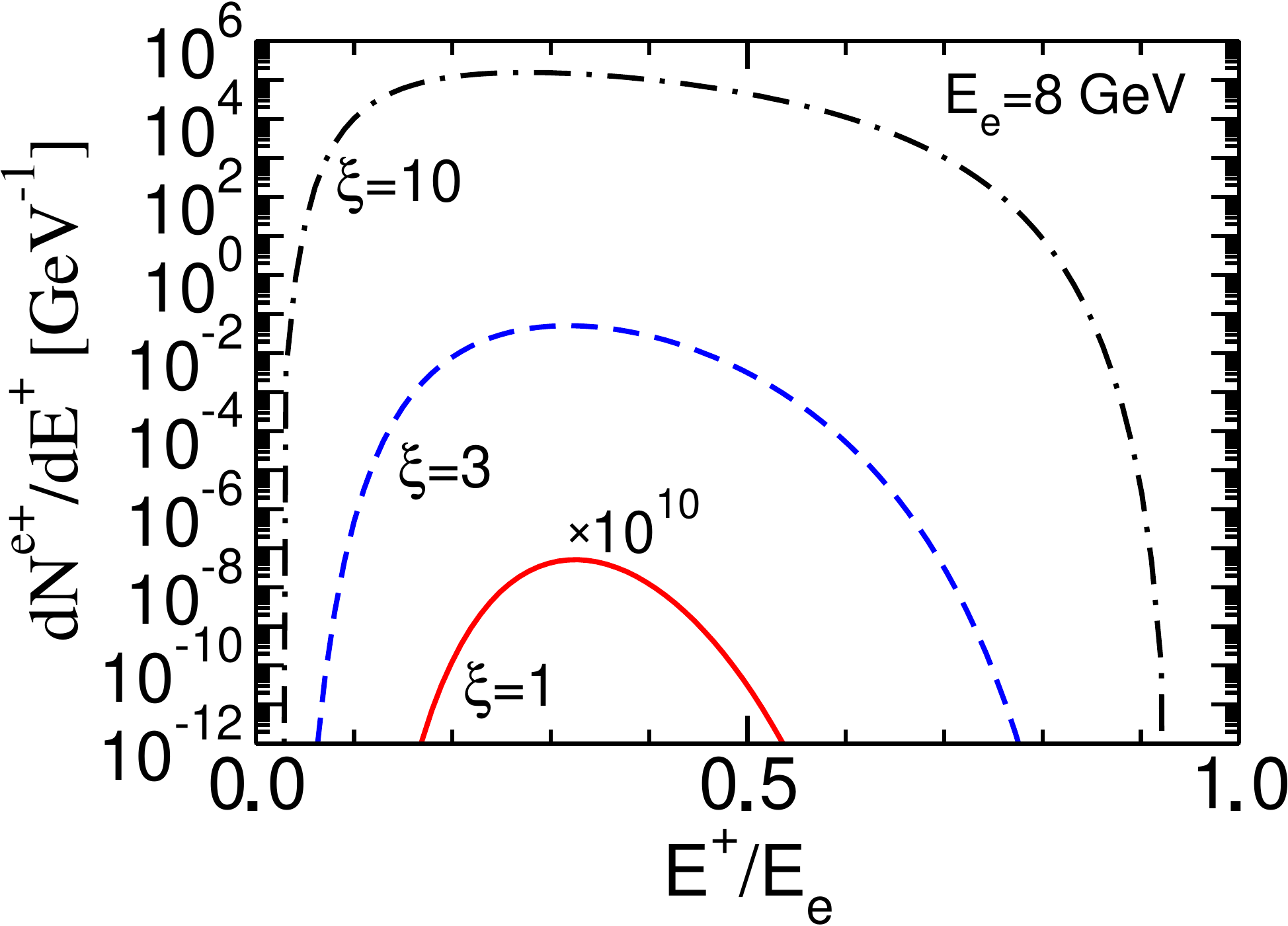}  
\includegraphics[width=0.66\columnwidth]{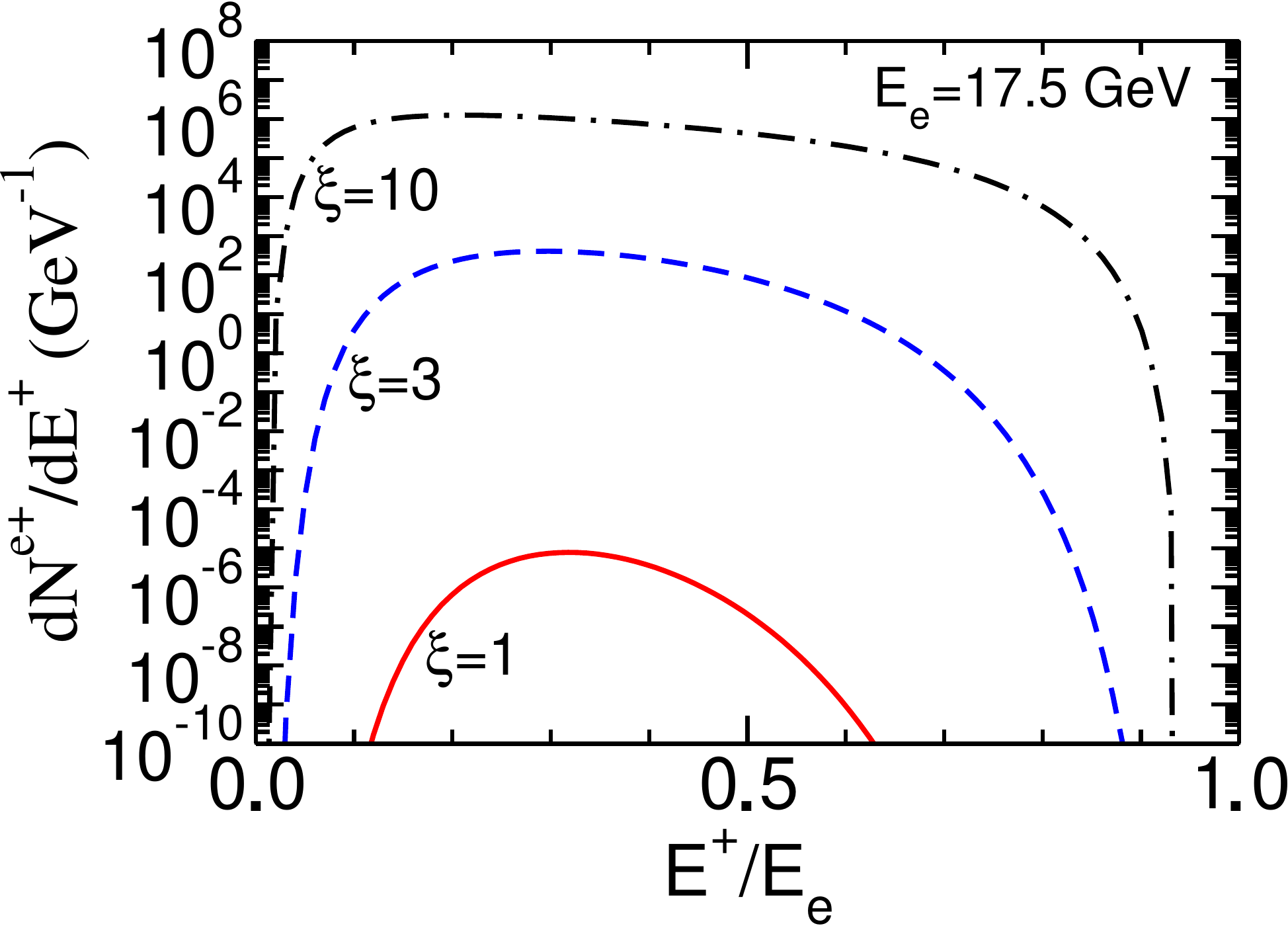} 
\includegraphics[width=0.66\columnwidth]{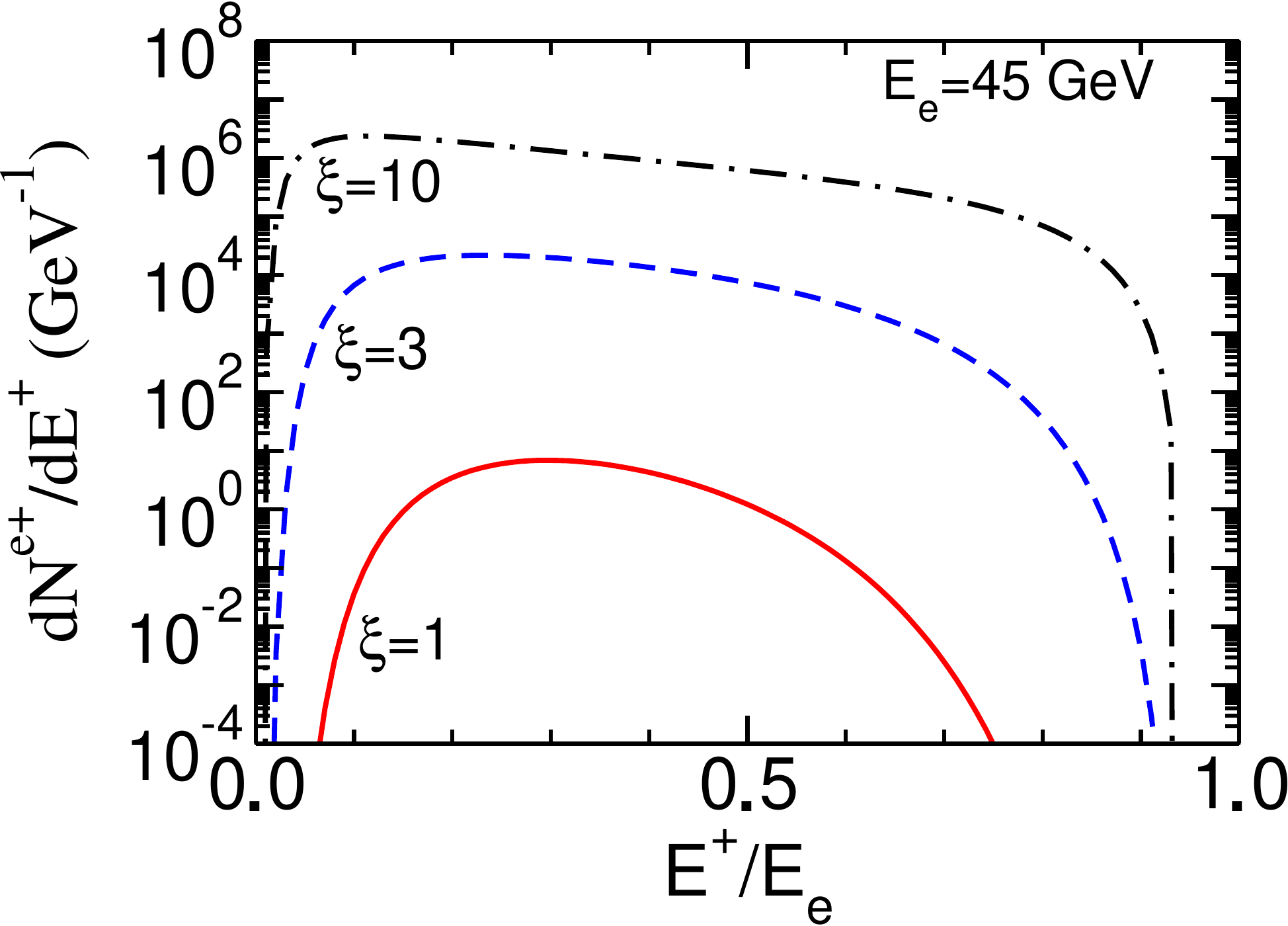}
\caption{(Color online)
The distribution of positrons $d N^{e^+}/dE^+$ as
a function of lab.\ energy $E^+$
at different values of $\xi = 1$ (red), 3 (blue dashed)
and 10 (black dot-dashed)
for electron beam energies $E_e = 8$~GeV (top panel),
17.5~GeV (middle panel) and 45~GeV (bottom panel) with $N_0=6\times 10^9$.
The harmonic structures of $d \Gamma_{n\ell C}/d \omega'$
seen in Fig.~\ref{Fig:01} at small values of $\omega'$ are irrelevant 
for $d N^{e^+}/dE^+$ since $d \Gamma_{n\ell BW}/d \omega'$
(see Fig.~\ref{Fig:021}) is exceedingly small there. 
\label{Fig:03}}
\end{figure}

\begin{figure}[t!]
\includegraphics[width=0.77\columnwidth]{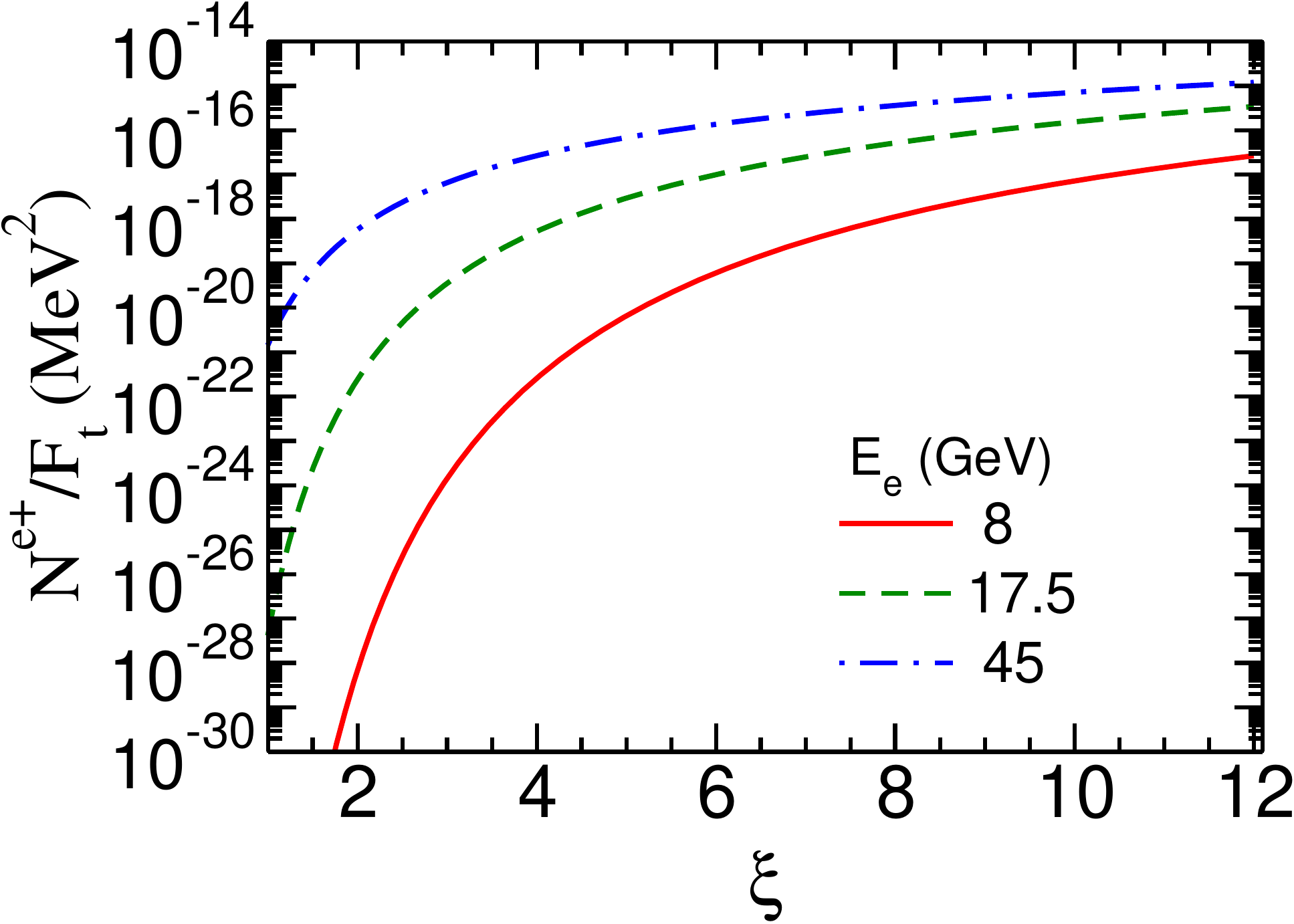}  
\includegraphics[width=0.77\columnwidth]{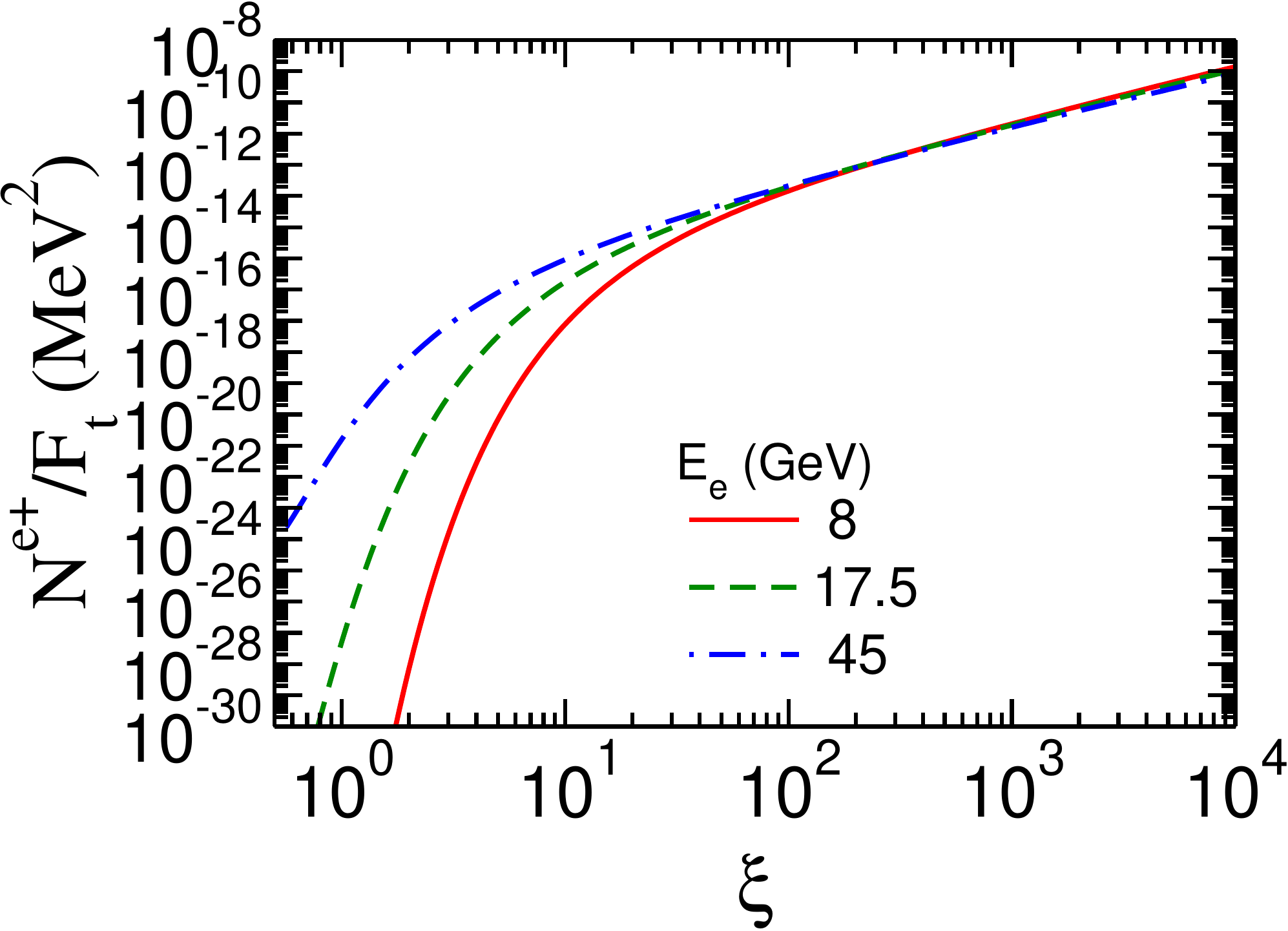}
\caption{(Color online)
The scaled number of positrons $N^{e^+} / F_t$ as
a function of laser intensity parameter $\xi$ for electron beam energies
$E_e = 8$~GeV (solid red), 17.5~GeV (dashed green) and 45~GeV (dot-dashed blue).
Top panel: for intensity parameter $\xi$ in the LUXE - FACET-II (E-320) range.
Bottom panel: for a large range of $\xi$. 
Note the value of the above given factor
%$F_t \approx \left( \frac{N_0^e}{10^{10}}\right)
%\left( \frac{1 \, \mbox{eV}}{\%omega}\right)^2 \, 
%2 \times 10^{23}$~MeV${}^{-2}$.
$F_t\,\omega^2\approx 20\,N_0$.
\label{Fig:04}}
\end{figure}

Our estimate of the positron distribution $d N^{e^+}/dE^+$
is exhibited in Fig.~\ref{Fig:03}.
For n$\ell$Co and n$\ell$BW vertices we us Eqs.(\ref{nlCo_approx})
and (\ref{nlBW},\,\ref{nlBW_approx2}), respectively.
The positron spectra display asymmetric bump-like distributions
with the maximum
shifted to smaller energies, i.e.\ to $E^+ < E_e/2$.
At large beam-electron energy the bump like distributions
become flatter and slightly inclined. The available phase
space, $0 < E^+ < E_e$,
becomes more and more uncovered with increasing values of $\xi$. 
Note the enormous sensitivity against variations of the laser intensity $\xi$. 

To quantify the $\xi$ dependence we display in Fig.~\ref{Fig:04}
the scaled positron number for three electron beam energies.
To convert to absolute numbers one has to multiply by the
beam electron number per bunch and adjust $T_C$ and $T_{BW}$
in Eq.~(\ref{E-distribution}) which determine the factor $F_t$
in Eq.~(\ref{E-distribution-final}). The top panel is for the laser intensity
range to be uncovered by LUXE - E-320. Using the above formalism also for
ultra-high intensities, the scaled positron number converge at $\xi > 50$
to a unique value irrespective of the electron beam energy $E_e$. Such a 
behavior is consistent with the stabilization phenomenon 
discussed in \cite{Kaminski:2006xlq}.

\begin{figure}[t!]
\includegraphics[width=0.77\columnwidth]{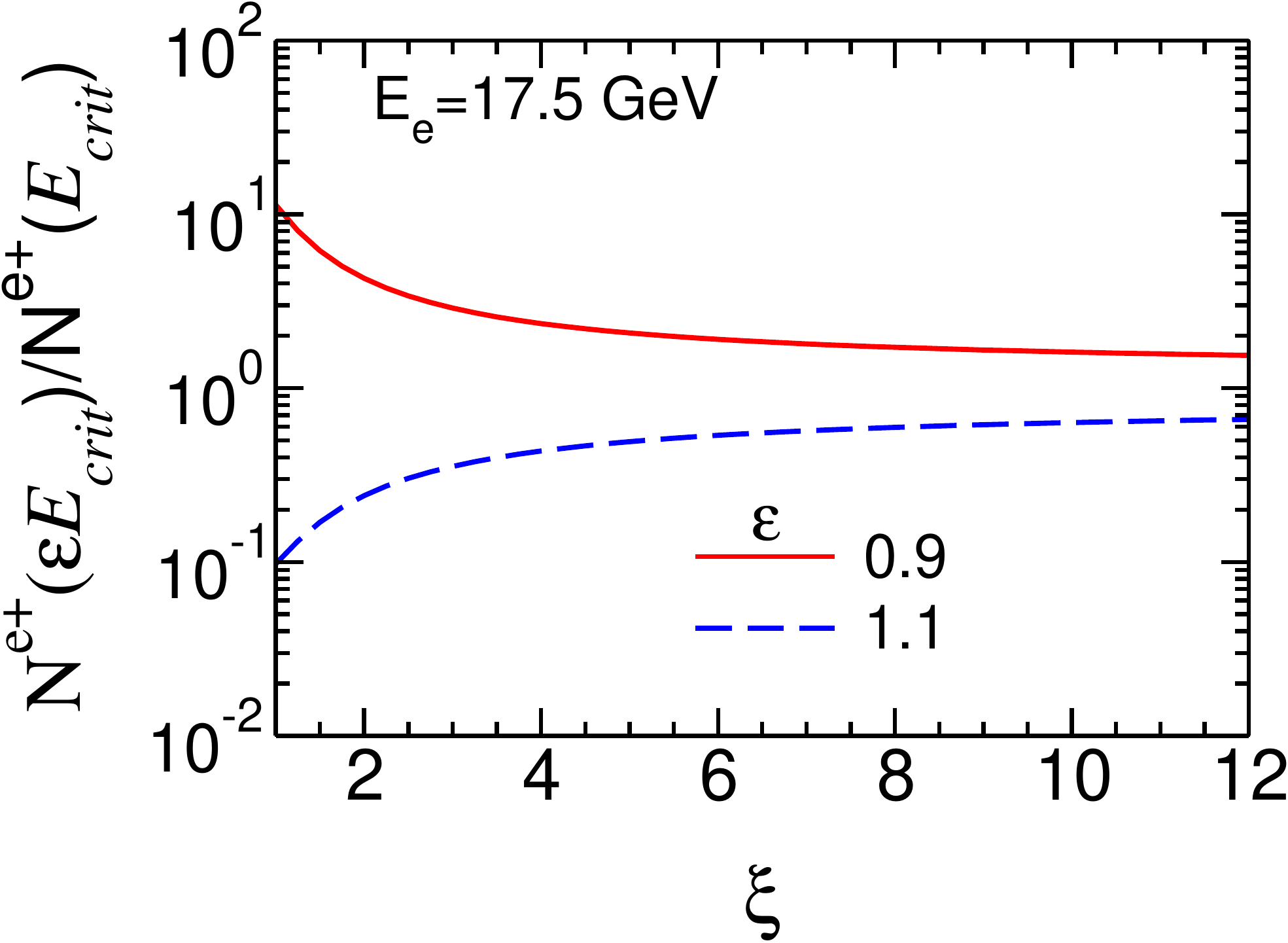}
\caption{(Color online)
The ratio $N^{e^+} (\epsilon \, {\cal E}_{crit}) / N^{e^+} ({\cal E}_{crit})$ as a function of $\xi$
for $E_e = 17.5$~GeV. Upper (lower) curve: $\epsilon = 0.9$ (1.1). 
Both curves approach unity for large values of $\xi$. 
\label{Fig:Ecrit}}
\end{figure}

The laser intensity parameter may be noted as $\xi = \frac{m}{\omega} \frac{{\cal E}}{{\cal E}_{crit}}$
with the (critical) Sauter-Schwinger electric field strength 
${\cal E}_{crit} = m^2/\vert e \vert \approx 1.3 \times 10^{18}$~V/m.
According to \cite{Hartin:2018sha}, ``Measuring the Boiling Point of the Vacuum of Quantum Electrodynamics'',
one can perform a deconvolution of data to get an experimental access to ${\cal E}_{crit}$, supposed all other
parameter are under control. Given the strong $\xi$ dependence of the individual contributions
in n$\ell$C($\xi$) $\otimes$ n$\ell$BW($\xi)$, see Figs.~1 and 2,
one can analogously test the dependence 
of the complete two-step positron number $N^{e^+}$ 
on ${\cal E}_{crit}$.
A useful quantity is the ratio
$N^{e^+} (\epsilon {\cal E}_{crit}) / N^{e^+} ({\cal E}_{crit}) = N^{e^+} (\xi / \epsilon) / N^{e^+} (\xi)$
as a function of $\xi$ 
with all other kinematic quantities fixed, see Fig.~\ref{Fig:Ecrit}.
Due to the exponential suppression of both the nonlinear Compton process \cite{HernandezAcosta:2020agu}
and the nonlinear Breit-Wheeler process \cite{Ritus85,Reiss:1971wf}, which is encoded in the
strong $\xi$ dependence, such a ratio exhibits in fact a stark sensitivity on variations of $E_{crit}$
on the 10\% level.

\section{Summary}\label{summary}

In summary we calculate, within a folding model n$\ell$C $\otimes$ n$\ell$BW
as approximation of the trident process, 
the energy distributions of positrons produced in collisions 
of ultra-relativistic electrons
with a high-intensity laser beam for different electron beam energies
and intensities of the laser beam. 
As a prerequisite we evaluate the gross shape 
(by ignoring the laser pule shape and thus not catching the complex interference pattern)  
of photon energy distributions from n$\ell$C, 
which are also experimentally accessible \cite{Fleck:2020opg}.
Our predictions are for parameters motivated by forthcoming experiments
with $\xi = 1 \cdots 10$, in particular at LUXE and FACET-II (E-320).
While the employed folding model is related to the two-step part of trident
in product approximation,
the interesting one-step contribution should manifest itself as deviation.
Our model may serve as reference easy to handle
in such a search for genuine strong-field QED effects.

\begin{figure}[!t]
\includegraphics[width=0.99\columnwidth]{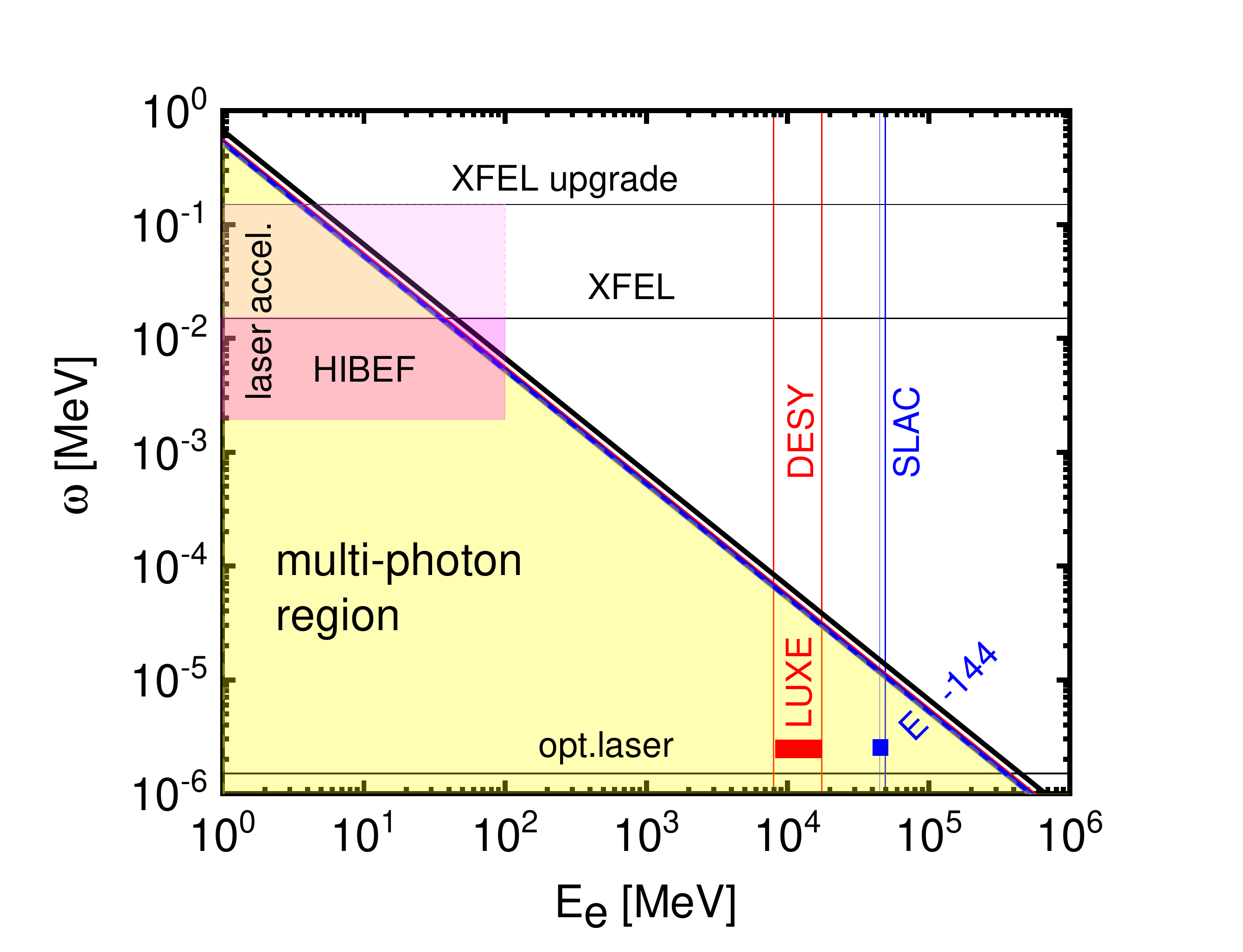}
\caption{(Color online)
Trident landscape over the $\omega$-$E_e$ plane.
The near-by diagonal lines depict the loci of $\sqrt{s} / m = 3.353$, 3.05, 3 and 2.95 (from top to bottom)
linear trident processes. The yellow region is below the threshold, i.e.\ accessible only by multi-photon
effects in the nonlinear trident or by  bandwidth effects in linear trident. 
Thin horizontal lines mark 1.5~eV, 15 keV and 150 keV as representatives of
optical lasers, XFEL and possible XFEL upgrade. Thin vertical lines bracket the energies of the Europaen XFEL
electron driver beam (DESY) and Stanford linear accelerator (SLAC). Boxes filled with color indicate the previous
E-144 (blue) and future LUXE (red) experiments as well as possible options at HIBEF (magenta)
when using laser accelerated
electrons with $E_e < 100$~MeV.      
\label{Fig:05}}
\end{figure}

In future developments one may envisage triple-vertex processes, e.g.\
irradiating laser-accelerated electrons in the laser focus by a high-energy XFEL beam,
$\mbox{XFEL} + (e^- + {\cal L})  \to {e^-}' + e^+ e^-$. This XFEL-induced trident
is a nonlinear virtual Compton process, which is accessible to the HIBEF collaboration
\cite{HIBEF}, in particular with an upgrade of the European XFEL.
(The corresponding theory of the nonlinear real Compton sub-process,
$\mbox{XFEL} + (e^- + {\cal L})  \to {e^-}' + \gamma'$,
has been already dealt with in collinear kinematics \cite{Seipt:2013hda}.)
An intermediate configuration could investigate the process
$\mbox{XFEL} + e_{{\cal L}}^-  \to {e^-}' + e^+ e^-$, again with laser-accelerated electrons ($e_{{\cal L}}^-$),
but outside the laser focus.
Since in the present setting the XFEL intensity parameter $\xi$ is small, $\xi_{\mbox{{\footnotesize XFEL}}} \ll 1$,
the multi-photon effects - which are decisive in section \ref{eL2pair} -
are negligible. The threshold electron energy is $E_e \approx 32$~MeV for 
$\omega_{\mbox{{\footnotesize XFEL}}} = 15$~keV.
Even below that value, pair production is enabled by bandwidth effects, see Appendix \ref{Appendix_A}.
If (by a yet unknown technology) the XFEL beam could be focused to achieve 
$\xi_{\mbox{{\footnotesize XFEL}}} \gtrsim 1$,
a new domain of testing strong-field QED effects would be opened up, see 
the magenta region in Fig.~\ref{Fig:05}.

%\newpage

\begin{appendix}

\section{Bandwidth effects in linear trident}\label{Appendix_A}

Turning to boost invariant quantities we describe the linearly polarized laser pulse 
by the electromagnetic four-potential
$A = (A^0, \vec A)$ in axial gauge, $A^0 = 0$, where
$\vec A = f (\phi) \, \vec a_x \, \cos \phi$
and  the envelope function reads 
$f (\phi) = \cos^2 \left( \frac{\pi \phi}{2 \Delta \phi} \right)
\sqcap (\phi, 2 \Delta \phi)$, 
i.e.\ the number of laser-field oscillations within the pulse is 
$N = \Delta \phi / \pi$. 
The invariant laser phase is $\phi = k \cdot x$.
Due to the finite pulse duration bandwidth effects occur, 
since the power spectrum of the laser has frequency components
$\lessgtr \omega$, thus enabling the sub-threshold pair production in the process
$\mbox{XFEL} + e^-  \to {e^-}' + e^+ e^-$.
The here assumed laser intensity parameter $\xi = e^2 {\vec a \,}^2 /m^2 \ll 1$
is too small to allow for genuine multi-photon effects.

\begin{widetext}

\begin{figure*}[h!]
  \includegraphics[width=0.77\columnwidth]{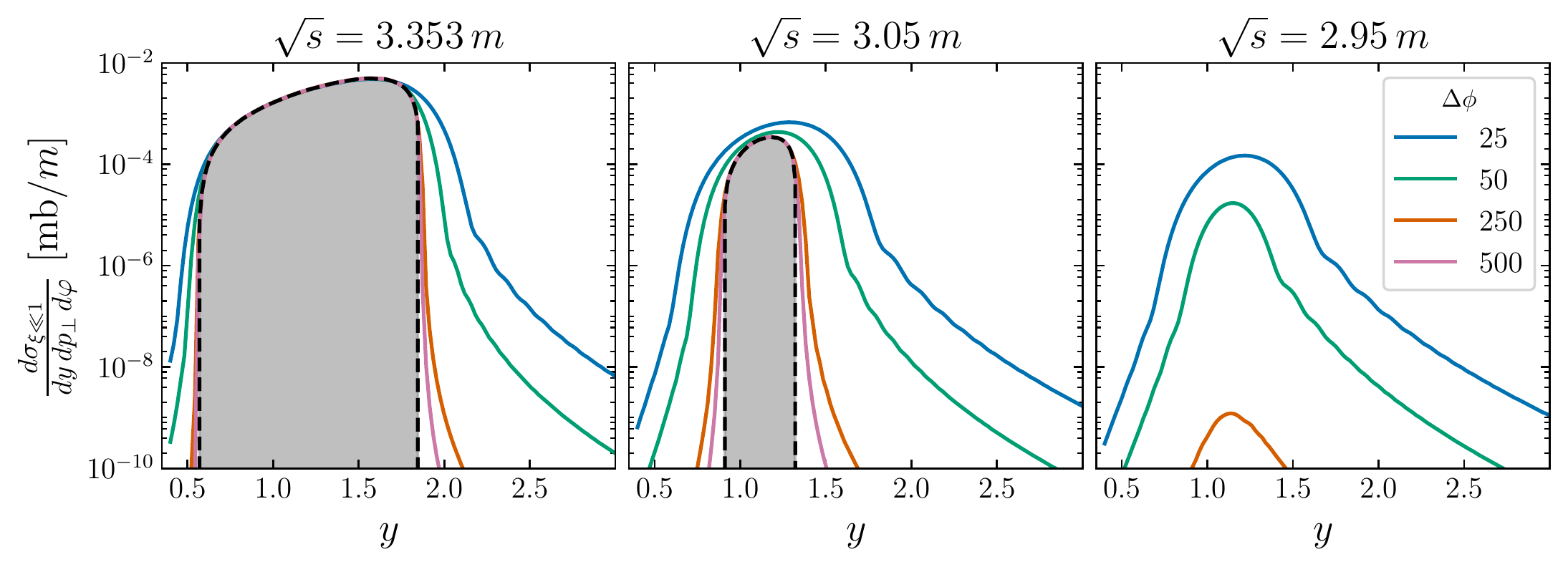}
%\vspace*{-3mm}
\caption{(Color online) 
Inclusive differential positron cross section $d^3 \sigma / d y \, d p_\perp \, d \varphi$ for
$\sqrt{s} /m  = 3.353$ (left panel), 3.05 (middle panel) and 2.95 (right panel)
as a function of rapidity $y$
for $p_\perp / m = 0.15$ and $\varphi = 0$. The pulse duration parameter is
$\Delta \phi = 25$ (blue), 50 (green), 250 (red) and 500 (magenta).
The black dashed curves limiting the gray region in left and middle panels 
depict the case of a monochromatic
beam, i.e.\ $\Delta \phi \to \infty$.
Details of the employed formalism and the kinematic relations can be found
in \cite{HernandezAcosta:2019vok,UHA}. 
\label{Fig:A1}}
\end{figure*}
\end{widetext}

In Fig.~\ref{Fig:A1} the inclusive differential positron cross section
$d^3 \sigma / d y \, d p_\perp \, d \varphi$ is exhibited, where
$y$, $p_\perp$ and $\varphi$ denote the rapidity, the transverse momentum and azimuthal angle
(w.r.t.\ to the polarization direction given by $\vec a_x$) of the produced positrons.  
The presented results are based on the weak-field evaluation of the l.h.s.\ Furry picture diagrams displayed
in the introduction.
We have selected three energies described by the Mandelstam variable $s = (p + k)^2$.
The threshold is given by $\sqrt{s} = 3 m$. The left and middle panels are above and near threshold,
while the right panel is below the threshold. The shorter the pulse, i.e.\ the smaller $\Delta \phi$,
the wider becomes the kinematically accessible region. In the limit of long pulses, $\Delta \phi \gg 1$,
the sub-threshold production is strongly suppressed. 

The figure applies to any position on the respective diagonal lines in Fig.~\ref{Fig:05} for given value of
$\sqrt{s}$ by shifting the
rapidity $y \to y + y_x$ with $y_x = \ln \sqrt{s} / m$.
The positron energy in the lab.\ frame, where $E_e = m \cosh y_e$, is 
$E^+ = \sqrt{m^2 + {p_\perp}^2} \cosh (y + y_x + y_e)$.

%\end{appendix}
%\begin{appendix}

\section{Testing EQ.(22) BY E-144 DATA}\label{Appendix_B}

%Testing Eq.~(\ref{E-distribution-final}) by E-144 data}

\begin{figure}[h!]
 \includegraphics[width=0.66\columnwidth]{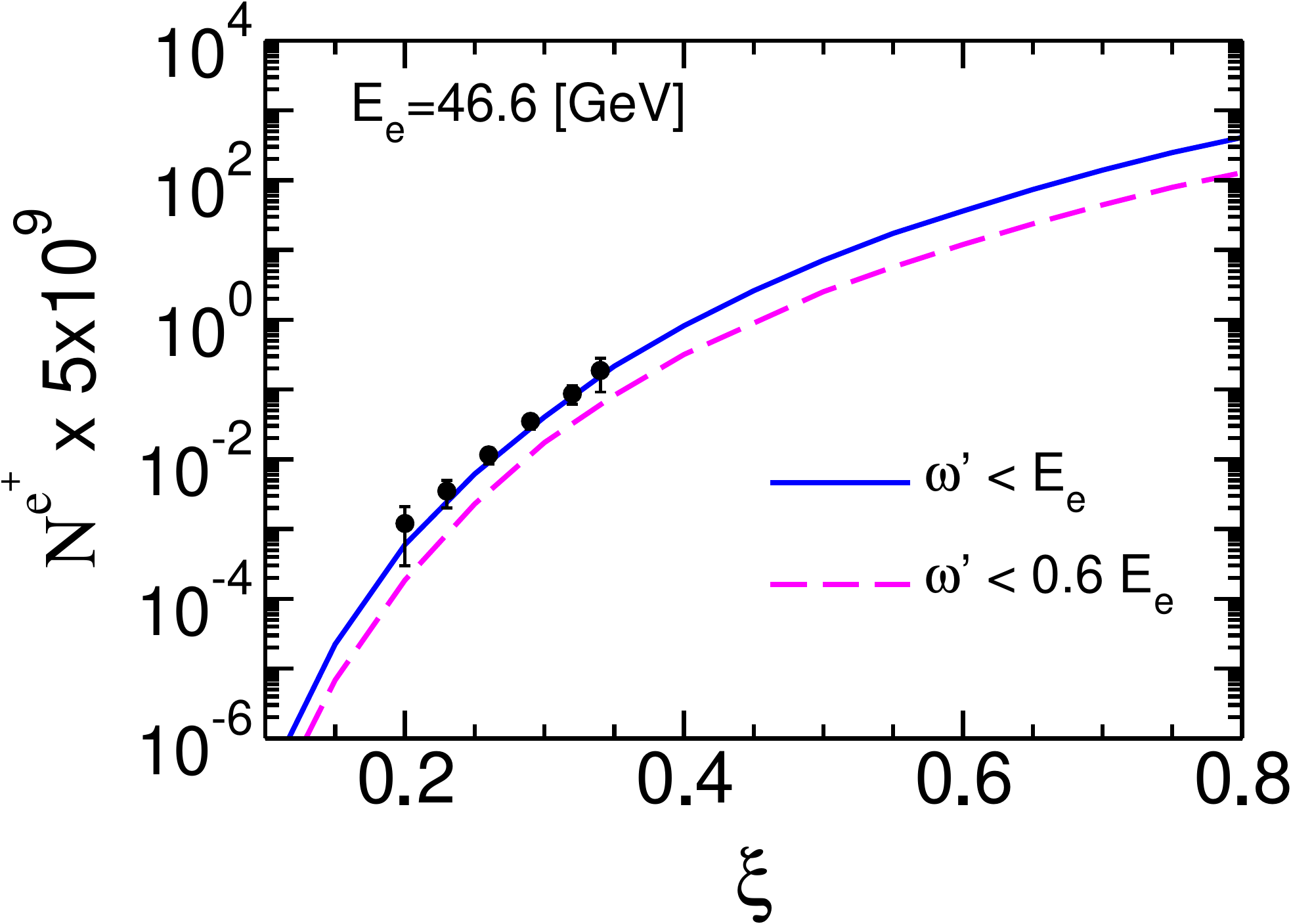}  
%\put(-200,120){[*] Testing Eq.~(\ref{E-distribution-final}) by E-144 data:}
\caption{(Color online) 
Positron number as a function of laser intensity parameter $\xi$
for $F_t = T_0^2 / 2$ with $T_0 = 2 \pi / \omega$ and $5 \times 10^9$ electrons per bunch
in Eq.~(\ref{E-distribution-final}).
The data are from \cite{Bamber:1999zt} (table~13). 
Accordingly the here employed laser frequency is $\omega = 2.3527$~eV.
\label{Fig:E-144}}
\end{figure}

\begin{figure}[h!]
\includegraphics[width=0.66\columnwidth]{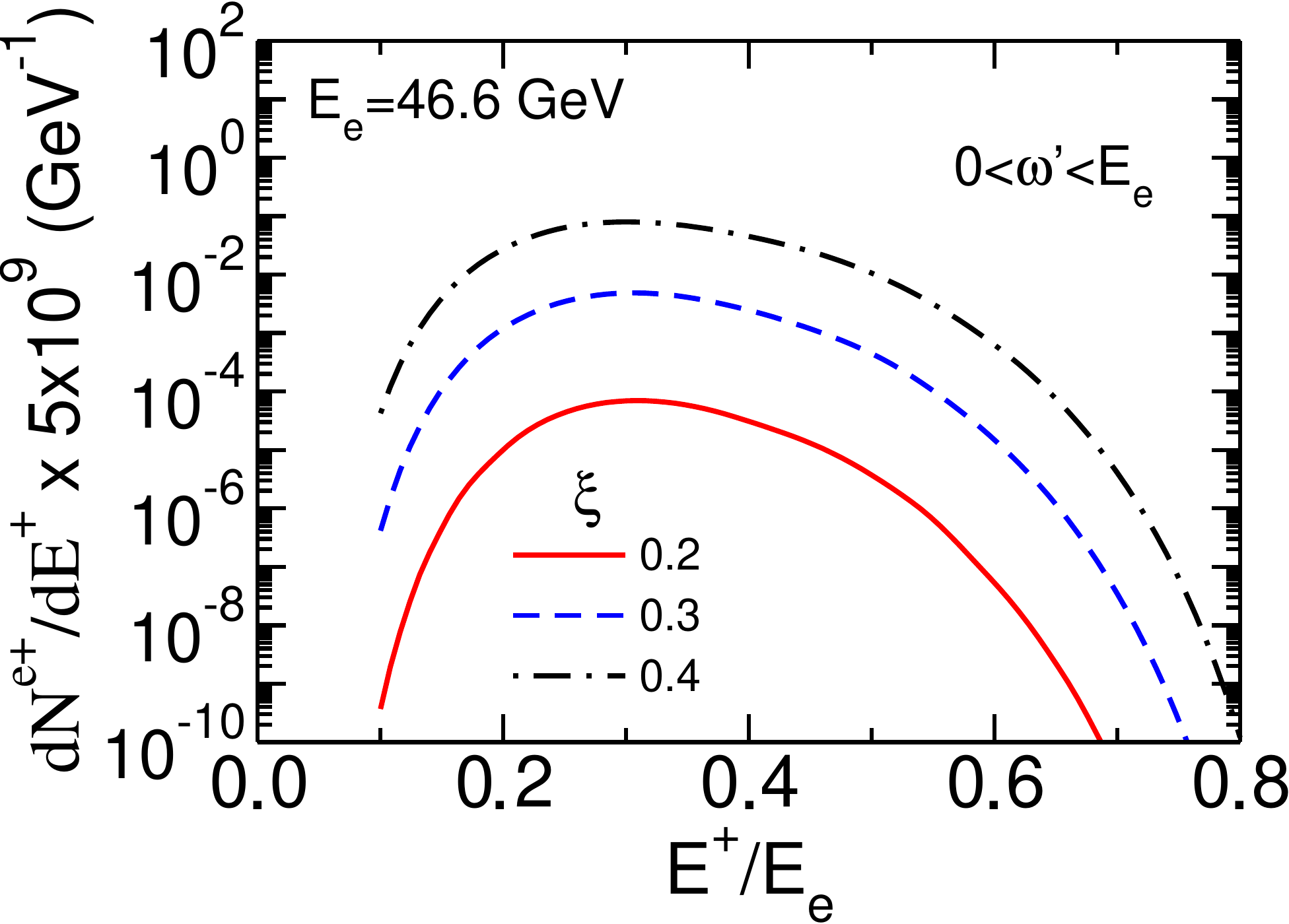}  
\caption{(Color online) 
Differential positron spectrum $dN^{e^+} / dE^+$ as a function of $E^+ /E_e$
with the same normalization as in Fig.~\ref{Fig:E-144} for $\xi = 0.2$ (red),
0.3 (blue dashed) and 0.4 (black dot-dashed). 
Despite the employed head-on geometry a value of $\xi = 0.246$ reproduces 
the peak height of \cite{Mackenroth:2018smh} (see figure 2 there)
when imposing the multiplicative factor
of $2 \times 10^4$ for the number of laser shots; our distribution is up shifted
by about 2~GeV and has a somewhat steeper l.h.s.\ flank.   
\label{Fig:E-144diff}}
\end{figure}

\begin{figure}[h!]
%\
 \includegraphics[width=0.66\columnwidth]{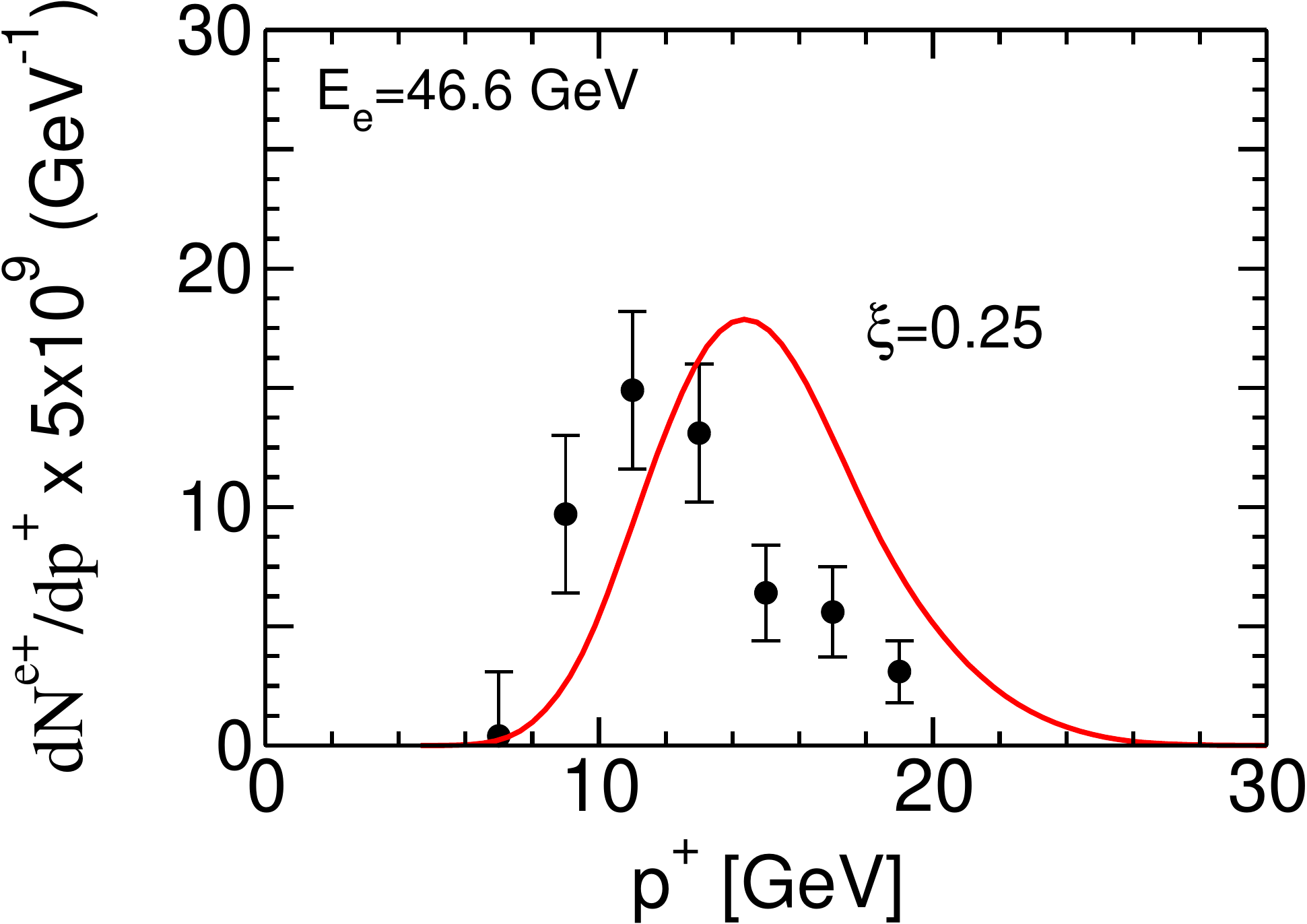} 
\caption{(Color online) 
Differential positron spectrum $dN^{e^+} / dp^+$ (red curve) as a function of 
the lab.\ positron momentum $p^+$
with the same normalization as in Figs.~\ref{Fig:E-144} and \ref{Fig:E-144diff} 
for $\xi = 0.25$ (see also \cite{Hu:2010ye})
and for 21,962 laser shots. 
Data from \cite{Bamber:1999zt} (table 12).
Including an inclination of $17^{\circ}$ between electron beam and laser beam
shifts the red curve to left.
\label{Fig:E-144diff_lin}}
\end{figure}

The choice $F_t = T_0^2 / 2$ with $T_0 = 2 \pi / \omega$
in Eq.~(\ref{E-distribution-final}) copes well with the $\xi$ dependence
of E-144 data (see Fig.~\ref{Fig:E-144}). Also the differential $E^+$ dependence
is sensitive to variations of $\xi$ (see Fig.~\ref{Fig:E-144diff}). 
An effective value of $\xi =0.25$ is consistent with the positron 
momentum distribution of E-144 (see Fig.~\ref{Fig:E-144diff_lin}).

\end{appendix}

\begin{acknowledgments}
  
The authors gratefully acknowledge communications with  G.~Torgrimsson,
A.~DiPiazza
and the longstanding collaboration with D.~Seipt.
One author (BK) thanks J.~Z.~Kaminski for explanations of the
stabilization phenomenon. 
The work is supported by R.~Sauerbrey and T.~E.~Cowan w.r.t.\ the study
of fundamental QED processes for HIBEF.
This work was partly funded by the Center for Advanced Systems
 Understanding (CASUS) that is financed by Germany’s
 Federal Ministry of Education and Research (BMBF) and by the
 Saxon Ministry for Science, Culture and Tourism (SMWK) with tax funds
 on the basis of the budget approved by the Saxon State Parliament.
\end{acknowledgments}

%\end{widetext}

%\newpage


\begin{thebibliography}{99}

%\cite{Loetstedt:2009zz}
\bibitem{Loetstedt:2009zz}
E.~Loetstedt and U.~D.~Jentschura,
``Correlated two-photon emission by transitions of Dirac-Volkov states in intense laser fields: QED predictions,''
Phys. Rev. A \textbf{80}, 053419 (2009)
%doi:10.1103/PhysRevA.80.053419
[arXiv:0911.4765 [quant-ph]].
%22 citations counted in INSPIRE as of 01 Jul 2021

%\cite{Seipt:2012tn}
\bibitem{Seipt:2012tn}
D.~Seipt and B.~K\"ampfer,
``Two-photon Compton process in pulsed intense laser fields,''
Phys. Rev. D \textbf{85}, 101701 (2012)
%doi:10.1103/PhysRevD.85.101701
[arXiv:1201.4045 [hep-ph]].
%66 citations counted in INSPIRE as of 01 Jul 2021

%\cite{Mackenroth:2012rb}
\bibitem{Mackenroth:2012rb}
F.~Mackenroth and A.~Di Piazza,
``Nonlinear Double Compton Scattering in the Ultrarelativistic Quantum Regime,''
Phys. Rev. Lett. \textbf{110}, no.7, 070402 (2013)
%doi:10.1103/PhysRevLett.110.070402
[arXiv:1208.3424 [hep-ph]].
%66 citations counted in INSPIRE as of 01 Jul 2021

%\cite{Ilderton:2010wr}
\bibitem{Ilderton:2010wr} 
  A.~Ilderton,
  ``Trident pair production in strong laser pulses,''
  Phys.\ Rev.\ Lett.\  {\bf 106}, 020404 (2011)
%  doi:10.1103/PhysRevLett.106.020404
  [arXiv:1011.4072 [hep-ph]].
  %%CITATION = doi:10.1103/PhysRevLett.106.020404;%%
  %96 citations counted in INSPIRE as of 17 Feb 2021

%\cite{King:2013osa}
\bibitem{King:2013osa}
B.~King and H.~Ruhl,
``Trident pair production in a constant crossed field,''
Phys. Rev. D \textbf{88}, no.1, 013005 (2013)
%doi:10.1103/PhysRevD.88.013005
[arXiv:1303.1356 [hep-ph]].
%59 citations counted in INSPIRE as of 15 Aug 2021

%\cite{King:2013zw}
\bibitem{King:2013zw}
B.~King, N.~Elkina and H.~Ruhl,
``Photon polarisation in electron-seeded pair-creation cascades,''
Phys. Rev. A \textbf{87}, 042117 (2013)
%doi:10.1103/PhysRevA.87.042117
[arXiv:1301.7001 [hep-ph]].
%68 citations counted in INSPIRE as of 15 Aug 2021

%\cite{Torgrimsson:2020wlz}
\bibitem{Torgrimsson:2020wlz} 
  G.~Torgrimsson,
  ``Nonlinear trident in the high-energy limit: Nonlocality, Coulomb field and resummations,''
  Phys.\ Rev.\ D {\bf 102}, no. 9, 096008 (2020)
 % doi:10.1103/PhysRevD.102.096008
  [arXiv:2007.08492 [hep-ph]].
  %%CITATION = doi:10.1103/PhysRevD.102.096008;%%
  %4 citations counted in INSPIRE as of 27 Dec 2020

%\cite{Dinu:2019pau}
\bibitem{Dinu:2019pau}
V.~Dinu and G.~Torgrimsson,
``Approximating higher-order nonlinear QED processes with first-order building blocks,''
Phys. Rev. D \textbf{102}, no.1, 016018 (2020)
%doi:10.1103/PhysRevD.102.016018
[arXiv:1912.11015 [hep-ph]].
%13 citations counted in INSPIRE as of 15 Aug 2021

%\cite{Dinu:2019wdw}
\bibitem{Dinu:2019wdw} 
  V.~Dinu and G.~Torgrimsson,
  ``Trident process in laser pulses,''
  Phys.\ Rev.\ D {\bf 101}, no. 5, 056017 (2020)
%  doi:10.1103/PhysRevD.101.056017
  [arXiv:1912.11017 [hep-ph]].
  %%CITATION = doi:10.1103/PhysRevD.101.056017;%%
  %7 citations counted in INSPIRE as of 27 Dec 2020

%\cite{Torgrimsson:2020gws}
\bibitem{Torgrimsson:2020gws}
G.~Torgrimsson,
``Loops and polarization in strong-field QED,''
New J. Phys. \textbf{23}, no.6, 065001 (2021)
% doi:10.1088/1367-2630/abf274
[arXiv:2012.12701 [hep-ph]].
%8 citations counted in INSPIRE as of 18 Oct 2021

%\cite{Dinu:2017uoj}
\bibitem{Dinu:2017uoj}
V.~Dinu and G.~Torgrimsson,
``Trident pair production in plane waves: Coherence, exchange, and spacetime inhomogeneity,''
Phys. Rev. D \textbf{97}, no.3, 036021 (2018)
% doi:10.1103/PhysRevD.97.036021
[arXiv:1711.04344 [hep-ph]].
%48 citations counted in INSPIRE as of 30 Oct 2021

%\cite{King:2018ibi}
\bibitem{King:2018ibi} 
  B.~King and A.~M.~Fedotov,
  ``Effect of interference on the trident process in a constant crossed field,''
  Phys.\ Rev.\ D {\bf 98}, no. 1, 016005 (2018)
%  doi:10.1103/PhysRevD.98.016005
  [arXiv:1801.07300 [hep-ph]].
  %%CITATION = doi:10.1103/PhysRevD.98.016005;%%
  %23 citations counted in INSPIRE as of 01 Jan 2021

%\cite{Mackenroth:2018smh}
\bibitem{Mackenroth:2018smh} 
  F.~Mackenroth and A.~Di Piazza,
  ``Nonlinear trident pair production in an arbitrary plane wave: a focus on the properties of the transition amplitude,''
  Phys.\ Rev.\ D {\bf 98}, no. 11, 116002 (2018)
%  doi:10.1103/PhysRevD.98.116002
  [arXiv:1805.01731 [hep-ph]].
  %%CITATION = doi:10.1103/PhysRevD.98.116002;%%
  %26 citations counted in INSPIRE as of 01 Jan 2021

%\cite{Hu:2010ye}
\bibitem{Hu:2010ye} 
  H.~Hu, C.~M\"uller and C.~H.~Keitel,
  ``Complete QED theory of multiphoton trident pair production in strong laser fields,''
  Phys.\ Rev.\ Lett.\  {\bf 105}, 080401 (2010)
%  doi:10.1103/PhysRevLett.105.080401
  [arXiv:1002.2596 [physics.atom-ph]].
  %%CITATION = doi:10.1103/PhysRevLett.105.080401;%%
  %104 citations counted in INSPIRE as of 17 Feb 2021

\bibitem{VIRitus}
V.~I.~Ritus,
``Vacuum polarization correction to elastic electron and muon scattering in an intense field
and pair electro- and muoproduction,"
Nucl.\ Phys.\ B {\bf 44}, 236 (1972) 

\bibitem{VNBaier}
V.~N.~Baier, V.~M.~Katkov, V.~M.~Strakhovenko,
``Higher-order effects in external fields: pair production by a particle,"
Sov.\ J.\ Nucl.\ Phys.\ {\bf 14}, 572 (1972) 

%\cite{Gonoskov:2014mda}
\bibitem{Gonoskov:2014mda} 
  A.~Gonoskov {\it et al.},
  ``Extended particle-in-cell schemes for physics in ultrastrong laser fields: Review and developments,''
  Phys.\ Rev.\ E {\bf 92}, no. 2, 023305 (2015)
%  doi:10.1103/PhysRevE.92.023305
  [arXiv:1412.6426 [physics.plasm-ph]].
  %%CITATION = doi:10.1103/PhysRevE.92.023305;%%
  %105 citations counted in INSPIRE as of 17 Feb 2021

%\cite{DiPiazza:2018bfu}
\bibitem{DiPiazza:2018bfu} 
  A.~Di Piazza, M.~Tamburini, S.~Meuren and C.~H.~Keitel,
  ``Improved local-constant-field approximation for strong-field QED codes,''
  Phys.\ Rev.\ A {\bf 99}, no. 2, 022125 (2019)
%  doi:10.1103/PhysRevA.99.022125
  [arXiv:1811.05834 [hep-ph]].
  %%CITATION = doi:10.1103/PhysRevA.99.022125;%%
  %39 citations counted in INSPIRE as of 27 Dec 2020

%\cite{Gonoskov:2021hwf}
\bibitem{Gonoskov:2021hwf}
A.~Gonoskov, T.~G.~Blackburn, M.~Marklund and S.~S.~Bulanov,
``Charged particle motion and radiation in strong electromagnetic fields,''
[arXiv:2107.02161 [physics.plasm-ph]].
%2 citations counted in INSPIRE as of 15 Aug 2021

%\cite{E_320}
\bibitem{E_320} 
S.~Meuren, ``Probing Strong-field QED at FACET-II (SLAC E-320) (2019)",
\begin{verbatim} 
https://conf.slac.stanford.edu/facet-2-2019/sites/
facet-2-2019.conf.slac.stanford.edu/files/
basic-page-docs/sfqed_2019.pdf.
\end{verbatim}

%\cite{Meuren:2020nbw}
\bibitem{Meuren:2020nbw} 
  S.~Meuren {\it et al.},
  ``On Seminal HEDP Research Opportunities Enabled by Colocating Multi-Petawatt Laser with High-Density Electron Beams,''
  arXiv:2002.10051 [physics.plasm-ph].
  %%CITATION = ARXIV:2002.10051;%%
  %8 citations counted in INSPIRE as of 02 Feb 2021

\bibitem{Abramowicz:2019gvx} 
  H.~Abramowicz {\it et al.},
  ``Letter of Intent for the LUXE Experiment,''
  arXiv:1909.00860 [physics.ins-det].
  %%CITATION = ARXIV:1909.00860;%%

%\cite{Abramowicz:2021zja}
\bibitem{Abramowicz:2021zja} 
  H.~Abramowicz {\it et al.},
  ``Conceptual Design Report for the LUXE Experiment,''
  arXiv:2102.02032 [hep-ex].
  %%CITATION = ARXIV:2102.02032;%%

%\cite{Salgado:2021fgt}
\bibitem{Salgado:2021fgt}
F.~C.~Salgado, N.~Cavanagh, M.~Tamburini, D.~W.~Storey, R.~Beyer, P.~H.~Bucksbaum, Z.~Chen, A.~Di Piazza, E.~Gerstmayr and Harsh, \textit{et al.}
``Single Particle Detection System for Strong-Field QED Experiments,''
[arXiv:2107.03697 [hep-ex]].
%0 citations counted in INSPIRE as of 09 Jul 2021

%\cite{Bamber:1999zt}
\bibitem{Bamber:1999zt} 
  C.~Bamber {\it et al.},
  ``Studies of nonlinear QED in collisions of 46.6-GeV electrons with intense laser pulses,''
  Phys.\ Rev.\ D {\bf 60}, 092004 (1999).
  %doi:10.1103/PhysRevD.60.092004
  %%CITATION = doi:10.1103/PhysRevD.60.092004;%%
  %236 citations counted in INSPIRE as of 26 Oct 2020

%\cite{Burke:1997ew}
\bibitem{Burke:1997ew} 
  D.~L.~Burke {\it et al.},
  ``Positron production in multi - photon light by light scattering,''
  Phys.\ Rev.\ Lett.\  {\bf 79}, 1626 (1997).
  %doi:10.1103/PhysRevLett.79.1626
  %%CITATION = doi:10.1103/PhysRevLett.79.1626;%%
  %413 citations counted in INSPIRE as of 26 Oct 2020

%\cite{Cole:2017zca}
\bibitem{Cole:2017zca} 
  J.~M.~Cole {\it et al.},
  ``Experimental evidence of radiation reaction in the collision of a high-intensity laser pulse with a laser-wakefield accelerated electron beam,''
  Phys.\ Rev.\ X {\bf 8}, no. 1, 011020 (2018)
%  doi:10.1103/PhysRevX.8.011020
  [arXiv:1707.06821 [physics.plasm-ph]].
  %%CITATION = doi:10.1103/PhysRevX.8.011020;%%
  %103 citations counted in INSPIRE as of 15 Jul 2020

%\cite{Poder:2018ifi}
\bibitem{Poder:2018ifi} 
  K.~Poder {\it et al.},
  ``Experimental Signatures of the Quantum Nature of Radiation Reaction in the Field of an Ultraintense Laser,''
  Phys.\ Rev.\ X {\bf 8}, no. 3, 031004 (2018)
%  doi:10.1103/PhysRevX.8.031004
  [arXiv:1709.01861 [physics.plasm-ph]].
  %%CITATION = doi:10.1103/PhysRevX.8.031004;%%
  %99 citations counted in INSPIRE as of 15 Jul 2020

\bibitem{Elkina:2010up} 
  N.~V.~Elkina  {\it et al.}, 
%A.~M.~Fedotov, I.~Y.~Kostyukov, M.~V.~Legkov,  N.~B.~Narozhny, E.~N.~Nerush and H.~Ruhl,
  ``QED cascades induced by circularly polarized laser fields,''
  Phys.\ Rev.\ ST Accel.\ Beams {\bf 14}, 054401 (2011),
%  doi:10.1103/PhysRevSTAB.14.054401
  [arXiv:1010.4528 [hep-ph]].

%\cite{Seipt:2020uxv}
\bibitem{Seipt:2020uxv}
D.~Seipt, C.~P.~Ridgers, D.~Del Sorbo and A.~G.~R.~Thomas,
``Polarized QED cascades,''
New J. Phys. \textbf{23}, no.5, 053025 (2021)
%doi:10.1088/1367-2630/abf584
[arXiv:2010.04078 [hep-ph]].
%6 citations counted in INSPIRE as of 15 Aug 2021

%\cite{King:2019igt}
\bibitem{King:2019igt}
B.~King,
``Uniform locally constant field approximation for photon-seeded pair production,''
Phys. Rev. A \textbf{101}, no.4, 042508 (2020)
% doi:10.1103/PhysRevA.101.042508
[arXiv:1908.06985 [hep-ph]].
%17 citations counted in INSPIRE as of 20 Oct 2021

%\cite{King:2020hsk}
\bibitem{King:2020hsk}
B.~King,
``Interference effects in nonlinear Compton scattering due to pulse envelope,''
Phys. Rev. D \textbf{103}, no.3, 036018 (2021)
%doi:10.1103/PhysRevD.103.036018
[arXiv:2012.05920 [hep-ph]].
%6 citations counted in INSPIRE as of 17 Oct 2021

 \bibitem{Acosta21} 
  U.~Hernandez Acosta, A.~I.~Titov and B.~Kämpfer,
  ``Rise and fall of laser-intensity effects
  in spectrally resolved Compton process,''
  New. J. Phys. {\bf 23}, 095008 (2021)/

\bibitem{Ritus85}
  V. I. Ritus,
 “Quantum effects of the interaction of elementary particles
with an intense electromagnetic field”.
  J. Sov. Laser Res. (United States), 6:5, 497 (1985).

%\cite{HernandezAcosta:2020agu}
\bibitem{HernandezAcosta:2020agu}
U.~Hernandez Acosta, A.~Otto, B.~K\"ampfer and A.~I.~Titov,
``Nonperturbative signatures of nonlinear Compton scattering,''
Phys. Rev. D \textbf{102}, no.11, 116016 (2020)
%doi:10.1103/PhysRevD.102.116016
[arXiv:2001.03986 [hep-ph]].
%4 citations counted in INSPIRE as of 08 Jul 2021

%\cite{Kampfer:2020cbx}
\bibitem{Kampfer:2020cbx}
B.~K\"ampfer and A.~I.~Titov,
``Impact of laser polarization on q-exponential photon tails in non-linear Compton scattering,''
Phys. Rev. A \textbf{103}, 033101 (2021)
%doi:10.1103/PhysRevA.103.033101
[arXiv:2012.07699 [hep-ph]].
%3 citations counted in INSPIRE as of 25 Aug 2021

%\cite{Tang:2021qht}
\bibitem{Tang:2021qht}
S.~Tang and B.~King,
``Pulse envelope effects in nonlinear Breit-Wheeler pair-creation,''
Phys. Rev. D {\bf 104} 096019 (2021).
%0 citations counted in INSPIRE as of 17 Oct 2021

%\cite{footnote_in_bibliography} 
\bibitem{footnote_in_bibliography} 
As alternative to the above approach which deploys rates,
one could use the convolution in (\ref{convolution}) without any
time ordering by turning to probabilities per pulse, analog to the 
``product approach" in \cite{King:2013osa}.

%\cite{Kaminski:2006xlq}
\bibitem{Kaminski:2006xlq}
J.~Z.~Kami\'nski, K.~Krajewska and F.~Ehlotzky,
``Monte Carlo analysis of electron-positron pair creation by powerful laser-ion impact,''
Phys. Rev. A \textbf{74}, no.3, 033402 (2006)
%doi:10.1103/PhysRevA.74.033402
%17 citations counted in INSPIRE as of 14 Aug 2021

%\cite{Hartin:2018sha}
\bibitem{Hartin:2018sha}
A.~Hartin, A.~Ringwald and N.~Tapia,
``Measuring the Boiling Point of the Vacuum of Quantum Electrodynamics,''
Phys. Rev. D \textbf{99}, no.3, 036008 (2019)
%doi:10.1103/PhysRevD.99.036008
[arXiv:1807.10670 [hep-ph]].
%24 citations counted in INSPIRE as of 16 Aug 2021

%\cite{Reiss:1971wf}
\bibitem{Reiss:1971wf}
H.~R.~Reiss,
``Production of electron pairs from a zero-mass state,''
Phys. Rev. Lett. \textbf{26}, 1072-1075 (1971)
%doi:10.1103/PhysRevLett.26.1072
%43 citations counted in INSPIRE as of 16 Aug 2021

%\cite{Fleck:2020opg}
\bibitem{Fleck:2020opg} 
  K.~Fleck, N.~Cavanagh and G.~Sarri,
  ``Conceptual Design of a High-flux Multi-GeV Gamma-ray Spectrometer,''
  Sci.\ Rep.\  {\bf 10}, no. 1, 9894 (2020).
  %doi:10.1038/s41598-020-66832-x
  %%CITATION = doi:10.1038/s41598-020-66832-x;%%

\bibitem{HIBEF}
HIBEF: Helmholtz International Beam Line for Extreme Fields, cf.\
\begin{verbatim}
https://www.hzdr.de/db/Cms?pOid=50566&pNid=694
\end{verbatim}

%\cite{Seipt:2013hda}
\bibitem{Seipt:2013hda}
D.~Seipt and B.~K\"ampfer,
``Laser assisted Compton scattering of X-ray photons,''
Phys. Rev. A \textbf{89}, no.2, 023433 (2014)
% doi:10.1103/PhysRevA.89.023433
[arXiv:1309.2092 [physics.atom-ph]].
%13 citations counted in INSPIRE as of 02 Jul 2021

%\cite{HernandezAcosta:2019vok}
\bibitem{HernandezAcosta:2019vok}
U.~Hernandez Acosta and B.~K\"ampfer,
``Laser pulse-length effects in trident pair production,''
Plasma Phys. Control. Fusion \textbf{61}, no.8, 084011 (2019)
% doi:10.1088/1361-6587/ab2b1e
[arXiv:1901.08860 [hep-ph]].
%13 citations counted in INSPIRE as of 02 Jul 2021

\bibitem{UHA}
U. Hernandez Acosta, 
``Pulsed perturbative QED: a study of trident pair production in pulsed laser fields,"
PhD thesis, TU Dresden, Germany (2020);
\begin{verbatim}
https://nbn-resolving.org/urn:nbn:de:
bsz:14-qucosa2-760352.
\end{verbatim}

\end{thebibliography}
\end{document}